\documentclass[a4paper,sort&compress, fleqn]{cas-sc}
\usepackage[numbers]{natbib}
\usepackage{textcomp}
\usepackage{url}
\usepackage{verbatim}
\usepackage{graphicx}
\usepackage{amssymb}
\usepackage{amsmath}
\usepackage{array,booktabs}
\usepackage{multirow}
\usepackage{bm}
\usepackage{bbding}
\usepackage{pifont}
\usepackage{makecell}
\usepackage{xcolor}
\usepackage{array}
\usepackage{tabularx}
\usepackage{listings}
\usepackage{hyperref}
\usepackage{subcaption}
\usepackage{multicol}

\usepackage[linesnumbered, ruled]{algorithm2e}
\usepackage[normalem]{ulem}
\usepackage{float}

\usepackage{xstring} 

\newcommand{\colorimpr}[1]{%
  \IfBeginWith{#1}{-}{%
    \textcolor{red!70!black}{#1\%}%
  }{%
    \textcolor{green!80!black}{#1\%}%
  }%
}
\newcommand{\dualimpr}[2]{\colorimpr{#1}/\colorimpr{#2}}

\def\tsc#1{\csdef{#1}{\textsc{\lowercase{#1}}\xspace}}
\tsc{WGM}
\tsc{QE}
\tsc{EP}
\tsc{PMS}
\tsc{BEC}
\tsc{DE}

\begin{document}
\shorttitle{}

\shortauthors{Rao et~al.}

\title [mode = title]{SAIL: Scene-aware Adaptive Iterative Learning for Long-Tail Trajectory Prediction in Autonomous Vehicles}

\author[1]{Bin Rao}
\credit{Conceptualization, Methodology, Experiment, Writing}

\author[2]{Haicheng Liao}
\credit{Methodology, Experiment, Writing}

\author[1]{Chengyue Wang}
\credit{Methodology, Experiment}

\author[3]{Keqiang Li}
\credit{Conceptualization, Methodology}

\author[4]{Zhenning Li}
\cormark[1]
\ead{zhenningli@um.edu.mo}
\credit{Conceptualization, Methodology, Funding}

\author[5]{Hai Yang}
\credit{Conceptualization, Methodology}

\affiliation[1]{organization={State Key Laboratory of Internet of Things for Smart City and Department of Civil and Environmental Engineering, University of Macau}, city={Macau SAR}, country={China}}

\affiliation[2]{organization={State Key Laboratory of Internet of Things for Smart City and Department of Computer and Information Science, University of Macau}, city={Macau SAR}, country={China}}

\affiliation[3]{organization={Department of Automotive Engineering, Tsinghua University}, city={Beijing}, country={China}}

\affiliation[4]{organization={State Key Laboratory of Internet of Things for Smart City and Departments of Civil and Environmental Engineering and Computer and Information Science, University of Macau}, city={Macau SAR}, country={China}}

\affiliation[5]{organization={Department of Civil and Environmental Engineering, The Hong Kong University of Science and Technology}, city={Hong Kong}, country={China}}

\cortext[cor1]{Corresponding author}

\begin{abstract}
Autonomous vehicles (AVs) rely on accurate trajectory prediction for safe navigation in diverse traffic environments, yet existing models struggle with long-tail scenarios—rare but safety-critical events characterized by abrupt maneuvers, high collision risks, and complex interactions. These challenges stem from data imbalance, inadequate definitions of long-tail trajectories, and suboptimal learning strategies that prioritize common behaviors over infrequent ones. To address this, we propose SAIL, a novel framework that systematically tackles the long-tail problem by first defining and modeling trajectories across three key attribute dimensions: prediction error, collision risk, and state complexity. Our approach then synergizes an attribute-guided augmentation and feature extraction process with a highly adaptive contrastive learning strategy. This strategy employs a continuous cosine momentum schedule, similarity-weighted hard-negative mining, and a dynamic pseudo-labeling mechanism based on evolving feature clustering. Furthermore, it incorporates a focusing mechanism to intensify learning on hard-positive samples within each identified class. This comprehensive design enables SAIL to excel at identifying and forecasting diverse and challenging long-tail events. Extensive evaluations on the nuScenes and ETH/UCY datasets demonstrate SAIL's superior performance, achieving up to 28.8\% reduction in prediction error on the hardest 1\% of long-tail samples compared to state-of-the-art baselines, while maintaining competitive accuracy across all scenarios. This framework advances reliable AV trajectory prediction in real-world, mixed-autonomy settings.
\end{abstract}


\maketitle
\begin{sloppypar}

\section{Introduction}\label{sec:intro}
The safe navigation of autonomous vehicles (AVs) in mixed-autonomy traffic environments hinges on accurately predicting the intentions and movements of dynamic traffic agents—vehicles, pedestrians, and cyclists \cite{liao2025minds, wang2025dynamics}. Recent advances in emerging deep learning technologies have driven a transition in autonomous driving (AD) research from traditional rule-based methods toward data-driven prediction models. Although these data-driven approaches have significantly improved prediction accuracy under common conditions, their performance suffers significant degradation in rare but safety-critical scenarios—a limitation rooted in the \textbf{long-tail distribution} of real-world trajectory data.

\begin{figure}[pos=t]
  \centering
  \includegraphics[width=0.6\linewidth]{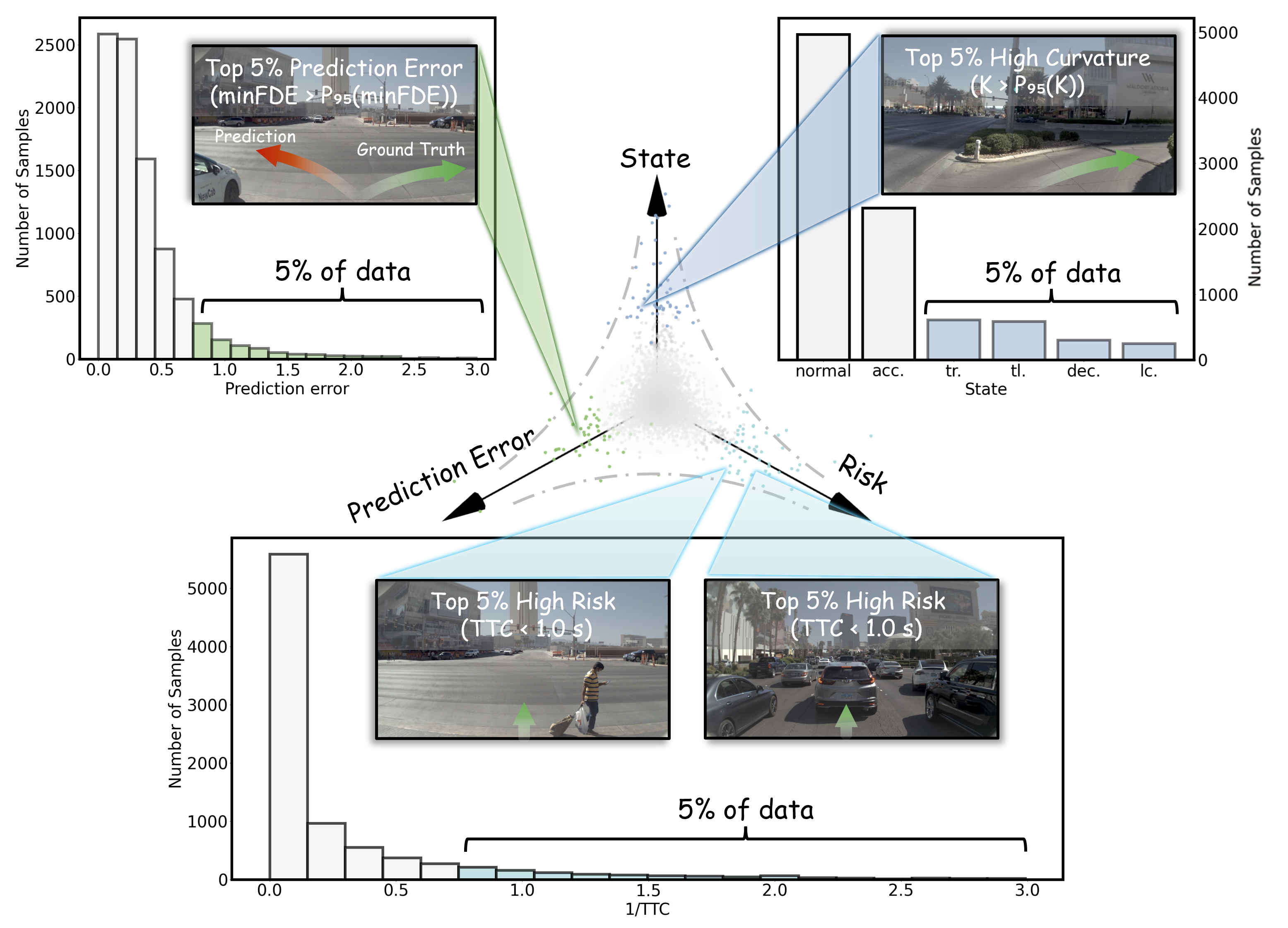} 
  \caption{Analyzing vehicle trajectories from the perspectives of Prediction Error, Risk (inverse time-to-collision, 1/TTC), and Vehicle State reveals their intrinsic long-tail nature. The top 5\% of data in each distribution corresponds to distinct real-world scenarios, which are often critical to ensuring the safe operation of autonomous vehicles.}
  \label{fig1} 
\end{figure}

In real-world traffic datasets, the majority of samples (the ``head'')  reflect common, predictable behaviors, such as smooth lane-following or gradual deceleration, which are straightforward for prediction models to learn. In contrast,  a small fraction of data (the ``tail'') represents complex, hazardous, and irregular situations involving abrupt lane changes, unpredictable pedestrian actions, or sudden obstacles, posing considerable challenges to existing models. These infrequent scenarios often involve high-risk interactions, posing significant safety threats. Current approaches, designed to minimize average-case errors, exhibit a natural bias towards common, simpler scenarios, often at the expense of marginalizing rarer and more complex cases. This misalignment originates from a core limitation: the scarcity of tail-class trajectories inherently restricts their influence during representation learning.

Addressing this limitation requires a principled framework to \textbf{define}, \textbf{identify}, and \textbf{adapt} to long-tail scenarios in alignment with real-world safety needs. However, existing literature still lacks a clear and consistent definition of ``long-tail trajectories''. Most studies tend to equate long-tail scenarios solely with data scarcity or prediction difficulty \cite{salzmann2020trajectron++,li2023graph-risk}, while neglecting other crucial domain-specific factors, such as collision risk severity, maneuver complexity, and the dynamics of agent interactions. As complex traffic systems illustrate, these concepts, although correlated, are not identical: a scenario may be statistically rare yet dynamically trivial to predict, or highly common yet extremely hazardous. Consequently, relying on a single perspective, such as statistical frequency alone or prediction loss alone, inevitably overlooks critical edge cases and hampers the proper prioritization of truly important long-tail scenarios. Second, the identification of tail samples tends to be highly model-dependent \cite{zhou2022long, lan2024hi-scl}, often relying on ad hoc prediction thresholds that vary across architectures. This approach overlooks the intrinsic characteristics of trajectories and compromises generalizability, as a scenario considered challenging by one model may be treated as trivial by another, thereby limiting robustness across diverse driving contexts. Furthermore, a critical limitation arises from the inability of learning strategies to effectively address the imbalance between common head samples and rare tail samples. Existing works \cite{salzmann2020trajectron++, liao2024bat-trans} to mitigate long-tail effects often employ conventional contrastive learning or augmentation strategies, which fail to adequately decouple the optimization objectives for head and tail classes, causing efforts to improve tail performance to degrade accuracy on common scenarios. Thus, there is a pressing need for learning frameworks capable of effectively modeling the diversity of tail-class scenarios while simultaneously maintaining robust performance across all trajectories, ultimately enhancing the safety and adaptability of autonomous driving systems.

As a recognized challenge in this field, long-tail prediction remains insufficiently explored in our community. Therefore, we focus on the imbalance inherent in trajectory prediction for autonomous vehicles and conduct an extended investigation. We first establish a rigorous definition of ``long-tail trajectories'' through a large-scale analysis of real-world driving data. Our analysis shows that long-tail phenomena arise not only from conventional class imbalance but also from dynamic operational constraints, such as low time-to-collision, abrupt lane changes, contextual complexity induced by occluded agents, and multimodal interactions. This multifaceted nature calls for a shift from simplistic frequency-based definitions to a criteria-driven framework. As illustrated in Figure \ref{fig1}, instead of relying on heuristic metrics, we propose a strictly data-driven three-way taxonomy that comprehensively captures the full spectrum of long-tail scenarios across three complementary information spaces: (1) \textit{Prediction Error} in the {model space}, which serves as a post hoc metric to capture contextual unpredictability and identify scenarios challenging for base predictive models; (2) \textit{Collision Risk} in the {relational space}, which is quantified through TTC-based metrics to capture spatiotemporal conflicts among multiple agents and highlight hazardous situations; and (3) \textit{State Complexity} in the {kinematic space}, which functions as a single-agent metric to capture maneuver nonlinearity and identify cases characterized by complex behaviors or partially observable environments. Together, this multidimensional and complementary formulation precisely isolates semantically rare and safety-critical events that may be overlooked by any single metric. Building on this multidimensional taxonomy, we introduce the Scene-aware Adaptive Iterative Learning framework, namely SAIL. SAIL is designed to reconcile strong performance in long-tail scenarios with robust overall accuracy. It begins with an attribute-guided data augmentation and feature extraction process that provides a rich and context-aware foundation for learning. The core of SAIL is a highly adaptive contrastive learning strategy that integrates unsupervised and supervised learning paradigms. Specifically, it employs a continuous cosine momentum schedule and similarity-weighted hard-negative mining to support robust initial representation learning. This process is followed by a dynamic pseudo-labeling mechanism based on evolving feature clustering, which provides high-quality supervision for a subsequent focused contrastive learning stage that strengthens learning on hard-positive samples. This comprehensive design ensures that SAIL is well equipped to address the diverse and challenging nature of long-tail trajectories.

Overall, our contributions are threefold:
\begin{itemize}
\item We take a step to systematically establish a comprehensive, data-driven methodology for identifying long-tail trajectory distributions based on three complementary dimensions: prediction error, collision risk, and state complexity. By bridging abstract mathematical metrics with concrete physical driving semantics, this multi-criteria approach provides a more robust and generalizable foundation for identifying rare, difficult, and safety-critical scenarios in real-world datasets.

\item We propose SAIL, a novel learning framework that introduces a highly adaptive, multi-stage contrastive learning strategy. This strategy synergizes attribute-guided learning with advanced techniques such as cosine momentum scheduling, weighted hard-negative mining, and a focusing mechanism for hard-positive samples, ensuring robust representation learning for imbalanced data.

\item Extensive evaluations on the nuScenes and ETH/UCY datasets demonstrate that SAIL delivers state-of-the-art (SOTA) performance in long-tail trajectory prediction. Our framework significantly enhances accuracy in rare and safety-critical scenes while maintaining leading overall prediction performance, confirming its adaptability and reliability for practical autonomous driving systems.
\end{itemize}

The structure of this paper is as follows: Section \ref{Related work} reviews the relevant literature. Section \ref{Methodology} details our model. Section \ref{Experiments} presents its performance on various datasets. Finally, Section \ref{Conclusion} concludes with a summary of the research.

\section{Related Work} \label{Related work}
\textbf{Trajectory Prediction in Autonomous Driving.}
Recent years have seen a rapid and transformative revolution in deep learning, with deep learning-based paradigms emerging as the primary solution to challenges in trajectory prediction tasks \cite{girgis2022latent, liao2025minds, shi2025rulenet}. Early studies employed computationally efficient methods based on classical kinematic and statistical models \cite{lin2000vehicle-yuce-1,wong2022view-yuce-2}, but these approaches face challenges in accurately modeling complex interactions and environmental uncertainties. The advent of deep learning has transformed the trajectory prediction landscape significantly. Data-centric methods, exemplified by VectorNet \cite{gao2020vectornet}, leverage rich contextual data to better capture spatial relationships. Architectures like Recurrent Neural Networks (RNNs) \cite{alahi2016social-lstm,huang2021bayonet-GRU,liao2024physics} further enabled effective modeling of temporal dependencies. Additionally, convolutional neural network-based approaches utilizing rasterized environmental maps \cite{gilles2021home, fan2025bidirectional} and social tensor representations \cite{deo2018convolutional, marchetti2024smemo, munir2025context} have successfully advanced spatial relationship extraction. 
More recently, transformer-based models, including Trajectron++ \cite{salzmann2020trajectron++}, HPP \cite{liu2025hybrid}, BAT \cite{liao2024bat-trans}, HLTP \cite{liao2024cognitive}, DEMO \cite{wang2025dynamics}, and MFTraj \cite{liao2024mftraj}, have excelled in simultaneously modeling complex spatial-temporal dependencies, particularly effective in dense and highly interactive scenarios.
Moreover, cutting-edge research has begun integrating large language models (LLMs) and generative world models to enhance prediction models' comprehension, zero-shot reasoning capabilities, and physical dynamics simulation in complex traffic scenarios \cite{wang2025wake, lan2024traj, liao2025cot, min2024driveworld}.


\textbf{Long-Tail Problem in Trajectory Prediction.}
A significant challenge in data-driven trajectory prediction is the long-tail distribution of driving behaviors \cite{zhou2022long}. To address this bottleneck, we summarize existing research and categorize current approaches into three distinct perspectives: (1) \textit{Prediction difficulty and representation learning}. Works in this paradigm define tail scenarios through training dynamics or high prediction errors \cite{makansi2021on-exposing, wang2023fend, zhang2024tract}. To mitigate overfitting on majority classes and extract robust representations for these ``hard'' minority samples, contrastive learning has emerged as a dominant technique \cite{chen2020simple-clr, yang2024dynamic}. Both unsupervised \cite{he2020momentum-moco} and supervised \cite{khosla2020supervised-cl, xuan2024decoupled} contrastive frameworks enhance feature separation; however, many studies fail to adequately decouple optimization objectives, inadvertently reducing accuracy on common patterns while improving tail performance. (2) \textit{Kinematic feature distribution frequency}. This perspective focuses on the statistical imbalance in the data count of underlying vehicle kinematics. Datasets are predominantly populated with common "head" maneuvers, such as stable lane-keeping, while rare "tail" events involving highly non-linear dynamics, such as abrupt lane changes, hard braking, or extreme yaw rates, remain exceedingly sparse \cite{shi2021improved, wang2025multi}. Beyond traditional data resampling \cite{han2005borderline-resample} or loss re-weighting \cite{ross2017focal-loss}, recent research has employed generative active learning via controllable diffusion models \cite{park2025generative} and LLM-driven frameworks like AGENTS-LLM \cite{yao2025agents} and Trajectory-LLM \cite{yang2025trajectory} to synthesize realistic, challenging traffic scenarios, thereby augmenting the training distribution to mitigate data scarcity. (3) \textit{Safety aspect and uncertainty}. This dimension targets worst-case scenarios, high-risk multi-agent interactions, and extreme corner cases regardless of their statistical occurrence \cite{li2023graph-risk, thuremella2024risk}. For instance, ensemble networks are frequently utilized to estimate predictive uncertainty from insufficient data to facilitate worst-case planning \cite{zhou2022long}. To better understand these complex edge cases, recent works have tokenized driving scenes to address long-tail events \cite{tian2024tokenize} and integrated Large Foundation Models with world models to perform common-sense reasoning for safety-critical corner cases \cite{liao2026addressing, lan2024traj}.

Despite the individual merits of these three perspectives, existing methods primarily treat them in isolation, resulting in a fragmented understanding of the long-tail distribution. Such single-dimensional criteria fail to capture the full spectrum of edge cases, as they cannot account for complex scenarios where prediction difficulty, spatial-temporal risk, and kinematic rarity intersect unexpectedly. To bridge this gap, our work distinguishes itself by unifying all three dimensions into a comprehensive, multi-criteria taxonomy. Furthermore, to address the representation entanglement seen in prior works, we propose a highly adaptive, dual-layer contrastive framework designed to decouple and prioritize underrepresented trajectory patterns across these complementary dimensions, improving safety-critical prediction accuracy without sacrificing overall performance.


\begin{figure*}[pos=t]
  \centering
\includegraphics[width=\textwidth]{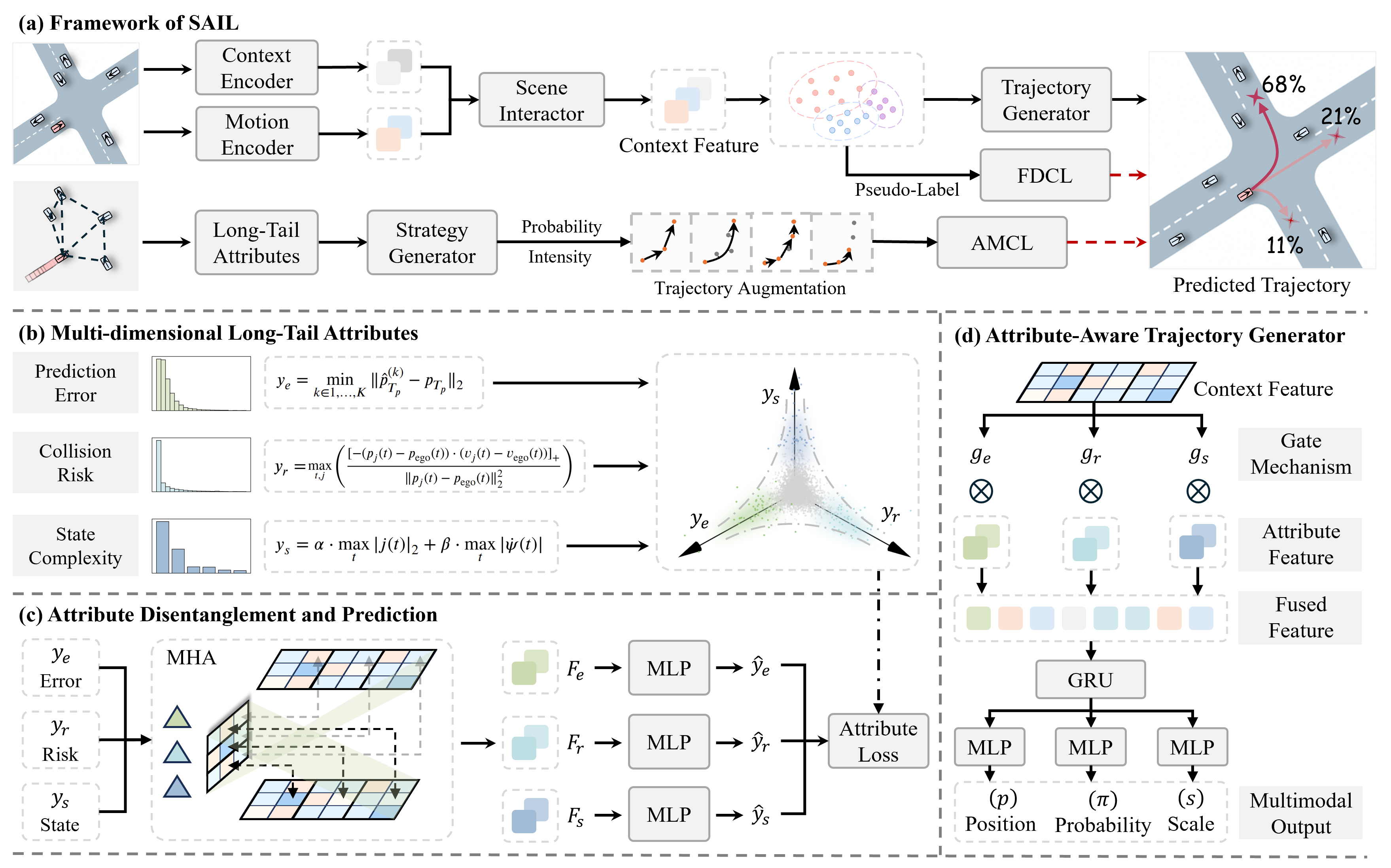} 
  \caption{The overall architecture of our proposed SAIL framework. The framework takes historical trajectories and HD map data as input and processes them through a multi-stage pipeline, including the Scene Representation Learning module and the Attribute-aware Trajectory Generator, to output multiple future trajectories. Panels (b), (c), and (d) provide detailed views of our key components: the Multi-dimensional Long-Tail Attributes definition, the Attribute Disentanglement and Prediction module, and the Attribute-aware Trajectory Generator, respectively.}
  \label{fig2} 
\end{figure*}

\section{Methodology} \label{Methodology}
\subsection{Problem Formulation}

We define the task of trajectory prediction as a sequence-to-sequence problem. For clarity, the main notations used throughout this paper are summarized in Table \ref{tab:notation}. Consider a traffic scene with $A$ total agents (vehicles and pedestrians). For any target agent $a \in \{1, ..., A\}$, its observed historical state over a past time horizon $t_{o}$ is denoted by $H_a = \{ {s}_a^t \mid t \in [0, t_{o}] \}$, where each state ${s}_a^t$ encapsulates kinematic information such as the agent's position, velocity, and heading. The static environment, including road lanes and crosswalks, is represented as a set of map vectors $M = \{ {m}_n \mid n=1, ..., N_m \}$, where $N_m$ is the total number of map elements. The objective is to predict the target agent's future trajectory over a future prediction horizon $t_{p}$. This ground-truth path is a sequence of future positions: $T = \{ {c}_t \mid t \in (t_{o}, t_{o}+t_{p}] \}$, where ${c}_t = (x_t, y_t)$ represents the target agent's spatial coordinates at a future time $t$. Therefore, the core task is to learn a predictive model that maps the observed history $H_a$ and the map context $M$ to a set of $K$ potential future trajectories $\hat{T} = \{ \hat{T}^{(1)}, \dots, \hat{T}^{(K)} \}$, where each $\hat{T}^{(k)} = \{ \hat{c}^{(k)}_t \mid t \in (t_{o}, t_{o}+t_{p}] \}$, so that the most likely predicted mode closely approximates the ground-truth $T$.

\begin{table*}[htpb]
\centering
\caption{Summary of main notations used throughout this paper.}
\label{tab:notation}
\resizebox{0.98\textwidth}{!}{
\begin{tabular}{ll|ll}
\toprule
\textbf{Symbol} & \textbf{Description} & \textbf{Symbol} & \textbf{Description} \\
\midrule
$A$ & Total number of agents & $a$ & Index of the target agent \\
$t_o$ & Observation horizon & $t_p$ & Prediction horizon \\
$H_a=\{s_a^t\}$ & Full observed historical state sequence & $s_a^t$ & Kinematic state of agent $a$ at time $t$ \\
$M=\{m_n\}$ & Set of map vectors & $N_m$ & Number of map elements \\
$T = \{ {c}_t \mid t \in (t_{o}, t_{o}+t_{p}] \}$ & Ground-truth future trajectory & $c_t=(x_t,y_t)$ & Ground-truth coordinate at time $t$ \\
$\hat{T}=\{\hat{T}^{(k)}\}_{k=1}^K$ & Set of predicted future trajectories & $K$ & Number of predicted modes \\
\midrule
$y_e$ & Prediction Error attribute & $y_r$ & Collision Risk attribute \\
$y_s$ & State Complexity attribute & $\hat{c}_{t_p}^{(k)}$ & Final predicted position of mode $k$ \\
$c_j(t),\,v_j(t)$ & Position and velocity of agent $j$ & $c_{\text{target}}(t),\,v_{\text{target}}(t)$ & Position and velocity of the target agent \\
$j(t)$ & Jerk at time $t$ & $\dot{\psi}(t)$ & Yaw rate at time $t$ \\
$\alpha,\beta$ & Weights in the state complexity metric &  &  \\
\midrule
$S$ & Augmentation strategy vector & $\phi_{\text{aug}}$ & Dynamic strategy generator \\
$T_o=\{c_1,\dots,c_{t_o}\}$ & Observed trajectory coordinates for augmentation & $T'_o$ & Augmented trajectory sequence \\
$\epsilon_{rdp}$ & RDP simplification threshold & $\Delta=(\delta_x,\delta_y)$ & Spatial shift vector \\
$\epsilon_{shift}$ & Shift magnitude & $\epsilon_{\max}$ & Maximum allowed shift magnitude \\
$b_i$ & Binary mask variable & $\rho$ & Retention probability in Mask augmentation \\
$\gamma$ & Subset ratio in Subset augmentation &  &  \\
\midrule
$F_a=\{F_{\text{target}},F_{\text{neighbors}}\}$ & Agent features & $F_m$ & Map feature representation \\
$A_M$ & Adjacency matrix of the map graph & $H_{\text{mode}} \in \mathbb{R}^{K\times D}$ & Modal queries \\
$F_p$ & Positional encoding & $F_{\text{context}} \in \mathbb{R}^{K\times D}$ & Multimodal context feature \\
$F_{\text{scene}} \in \mathbb{R}^{D}$ & Aggregated scene-level feature & $F_e,F_r,F_s$ & Disentangled attribute features \\
$\hat{y}_e,\hat{y}_r,\hat{y}_s$ & Predicted attribute values & $\phi_{\text{MHA}}$ & Multi-head attention module \\
$\phi_{\text{MLP}_e},\phi_{\text{MLP}_r},\phi_{\text{MLP}_s}$ & Attribute prediction heads &  &  \\
\midrule
$m_b,m_f$ & Initial and final momentum coefficients & $E_{\max}$ & Total training epochs \\
$\theta_q,\theta_k$ & Parameters of query and momentum encoders & $Q$ & Negative sample queue in AMCL \\
$N_{neg}$ & Number of selected hard negatives & $s^+,\,s_i^-$ & Positive and hard-negative similarities \\
$w_i$ & Weight of the $i$-th hard negative & $\tau_a,\tau_w$ & Temperature parameters in AMCL \\
$L_{amcl}$ & AMCL loss &  &  \\
\midrule
$\mathcal{B}=\{f_i\}_{i=1}^N$ & Feature memory bank & $C$ & Number of clusters in EFC \\
$\mu_j$ & Cluster centroid & $PL_i^{(e)}$ & Pseudo-label of feature $f_i$ at epoch $e$ \\
$\mathcal{P}_i$ & Positive set sharing the same pseudo-label & $\mathcal{N}_i$ & Negative set from other pseudo-label classes \\
$w_r$ & Class-aware weight in FDCL & $w_f$ & Focusing weight in FDCL \\
$w_{fdcl}$ & Final positive-pair weight in FDCL & $\eta$ & Focusing hyperparameter in FDCL \\
$\tau_f$ & Temperature parameter in FDCL & $L_{fdcl}$ & FDCL loss \\
\midrule
$g$ & Attribute gating weights & $g_e,g_r,g_s$ & Gates for three attribute features \\
$F_{\text{gated}}$ & Gated attribute-aware feature & $F_{\text{fused}}$ & Final fused feature \\
$h_t$ & Hidden state of the GRU decoder & $h_0$ & Initial decoder hidden state \\
$z_t$ & Learnable decoder query at time step $t$ & $c_t^k,s_t^k,\pi^k$ & Predicted coordinate, scale, and mode probability \\
$L_{task}$ & Main trajectory prediction loss & $L_{target},L_{reg},L_{cls}$ & Task-loss components \\
$L_{attr}$ & Auxiliary attribute supervision loss & $\lambda_1,\lambda_2,\lambda_3$ & Loss weights \\
$L$ & Overall training objective &  &  \\
\bottomrule
\end{tabular}
}
\end{table*}

\subsection{Overall Framework} The overall framework of SAIL, illustrated in Figure \ref{fig2}, systematically addresses long-tail trajectory prediction through a multi-stage pipeline. The process initiates by extracting long-tail attributes from three dimensions. Guided by these attributes, our Attribute-Guided Trajectory Augmentation (AGTA) module employs an augmentation strategy generator to create targeted policies, producing a diverse set of augmented trajectory data. Both original and augmented trajectories are then processed via Scene Representation Learning and Attribute Disentanglement, yielding rich, attribute-aware feature representations. These features first enter the Adaptive Momentum Contrastive Learning (AMCL) module for robust unsupervised representation learning. Subsequently, our Evolving Feature Clustering (EFC) strategy generates dynamic pseudo-labels from these features, which in turn provide supervision for the Focused Decoupled Contrastive Learning (FDCL) module to further refine the feature space. Finally, the resulting discriminative features are passed to an Attribute-aware Trajectory Generator to predict a set of diverse future trajectories.

\subsection{Multi-dimensional Long-Tail Attributes}

Existing research often characterizes long-tail trajectories from a singular perspective, such as prediction difficulty, leading to a one-dimensional and incomplete understanding of the problem. To address this limitation, we propose a Multi-dimensional Attribute Framework that provides a comprehensive and fine-grained representation of long-tail scenarios. We identify three distinct yet complementary attributes: Prediction Error, Collision Risk, and State Complexity. These three dimensions systematically cover complementary information spaces. Specifically, Prediction Error serves as a post-hoc, model-level metric that captures the contextual unpredictability of a scenario. In parallel, Collision Risk and State Complexity serve as objective, model-agnostic physical metrics derived directly from ground-truth kinematics. The former quantifies multi-agent spatial-temporal conflicts, while the latter measures the intrinsic non-linearity of single-agent maneuvers. By integrating these model-dependent and physical perspectives, we provide a comprehensive and unbiased formulation of the long-tail distribution.

\textbf{Prediction Error ($y_e$).} This attribute quantifies the intrinsic predictability of a trajectory, identifying samples that deviate from common motion patterns. We define it as the Final Displacement Error (FDE) between the ground-truth future trajectory $T$ and predictions $\hat{T}$ from a pre-trained baseline model, for which we use the well-established Trajectron++ \cite{salzmann2020trajectron++}. To ensure robustness, we select the minimum FDE over $K$ predicted modalities.
\begin{equation}
y_{e} = \min_{k \in {1, \dots, K}} \| \hat{c}_{t_p}^{(k)} - c_{t_p}\|_2
\label{eq:pred_error}
\end{equation}
where $\hat{c}_{t_p}^{(k)}$ and $c_{t_p}$ are the final position of ground-truth and the $k$-th predicted trajectory, $\| \cdot \|_2$ denotes the $L_2$ norm.

\textbf{Collision Risk ($y_r$).} The distribution of risk in autonomous driving scenarios is inherently long-tailed, forming a "risk pyramid." The wide base of this pyramid consists of frequent, low-risk scenarios, while the narrow apex is composed of rare, high-risk, safety-critical events such as near-collisions. These high-risk, low-frequency events are a critical component of the long-tail problem, as they represent the most challenging and consequential situations for an autonomous system. We use the Inverse Time-to-Collision (InvTTC) to quantify this risk dimension, which is the reciprocal of the time remaining before two agents collide assuming they maintain their current velocities. The risk is defined as the maximum InvTTC observed between the target agent and any other agent $j$ over the entire trajectory.
\begin{equation}
y_r = \max_{t, j} \left( \frac{[ -({c}_j(t) - {c}_{\text{target}}(t)) \cdot ({v}_j(t) - {v}_{\text{target}}(t)) ]_+}{\| {c}_j(t) - {c}_{\text{target}}{(t)}  \|_2^2} \right)
\label{eq:collision_risk}
\end{equation}
where $[x]_+ = \max(0, x)$, ${c}(t)$ and ${v}(t)$ are the time-dependent position and velocity vectors, respectively. A high $y_r$ value thus signifies a significant and immediate collision risk, identifying a key sample from the tail of the risk distribution.

\textbf{State Complexity ($y_s$).} This attribute captures the kinematic complexity and non-linearity of an agent's motion. We formulate it as a weighted combination of the maximum jerk (rate of change of acceleration, ${j}$) and the maximum yaw rate ($\dot{\psi}$), which together describe both translational and rotational irregularities.
\begin{equation}
y_s = \alpha \cdot \max_t \left| {j}(t) \right|_2 + \beta \cdot \max_t \left| \dot{\psi}(t) \right|
\label{eq:state_complexity}
\end{equation}
where ${j}(t) = \frac{d{a}(t)}{dt}$ and $\dot{\psi}(t)$, weighted by $\alpha$ and $\beta$ respectively, are computed over the entire trajectory. This metric effectively identifies erratic or highly dynamic maneuvers characteristic of long-tail scenarios.

\subsection{Attribute-Guided Trajectory Augmentation}
Data augmentation is a cornerstone for improving model generalization, especially for imbalanced long-tail distributions. However, conventional augmentation methods apply transformations uniformly and stochastically, a strategy that is suboptimal for the structured nature of trajectory data. Such attribute-agnostic approaches risk applying irrelevant or even detrimental augmentations, for instance, over-simplifying an already simple trajectory or applying minor shifts to a scenario where collision risk is the dominant challenge.

To address this challenge, we introduce a novel Attribute-Guided Augmentation module. Instead of random selection, our approach learns to generate a tailored augmentation strategy for each trajectory based on its attribute vector $(y_e, y_r, y_s)$. This ensures that the augmentation is not only relevant but also maximally informative for improving the model's robustness against specific long-tail challenges. Our framework consists of two core components: a learnable Dynamic Strategy Generator and a set of Attribute-Specific Augmentation Functions.

\subsubsection{Dynamic Strategy Generator}
We propose a Dynamic Strategy Generator designed to learn a sophisticated mapping from the problem space, defined by trajectory attributes, to the solution space of data augmentation strategies. Rather than relying on handcrafted rules, this module automatically discovers the most effective augmentation method by analyzing the specific long-tail characteristics of a given trajectory. The underlying principle is that different types of long-tail scenarios necessitate different forms of data augmentation to be maximally effective. For instance, scenarios with high State Complexity may benefit most from the Simplify strategy to isolate the core motion pattern, whereas high Collision Risk scenarios are better addressed by the Shift strategy to create challenging, safety-critical variants. This generator is implemented as a lightweight Multi-Layer Perceptron ($\phi_{{aug}}$) that takes the 3-dimensional attribute vector as input. It outputs a policy vector containing the selection probabilities and intensity parameters for the augmentation functions.
\begin{equation}
S = \phi_{aug}(y_e, y_r, y_s)
\end{equation}
where ${S}$ contains the probability scores for applying Simplify, Shift, Mask, and Subset, respectively, and the normalized intensity parameters. For each sample, we select the single augmentation corresponding to the highest probability. This attribute-guided selection ensures a stable and targeted transformation.

\subsubsection{Augmentation Functions}

We design four augmentation functions to generate diverse yet semantically meaningful trajectory views under attribute-guided control. These augmented trajectories serve as additional contrastive views for representation learning, while the main forecasting backbone continues to operate on the original trajectory stream. Each function perturbs a different aspect of motion patterns, and the probability of applying each augmentation is dynamically determined by the strategy generator. Let the input sequence be the observed historical trajectory $T_{o} = \{c_1, c_2, \dots, c_{t_o}\}$, where $t_o$ is the observation length.

\textbf{(a) Simplify.} Long-tail trajectories, such as sharp turns or evasive maneuvers, are often defined by a few critical shape-defining points rather than a dense sequence. To help the model focus on these essential geometric features, we employ the Ramer-Douglas-Peucker (RDP) algorithm. The primary goal of this algorithm is to simplify the trajectory by filtering out minor positional jitters while preserving significant inflection points. This targeted simplification distills the core motion pattern of a rare maneuver, ensuring the model focuses on the key geometric characteristics rather than being distracted by trivial fluctuations. The algorithm produces a simplified trajectory $T'_{o}$:
\begin{equation}
 T'_{o} = \{c_1, \dots, c_k, \dots, c_{t_o}\} \quad 
\text{s.t.} \quad \text{dist}(c_i, \text{line}(c_{s}, c_{e})) \le \epsilon_{rdp} 
\end{equation}
where $c_s$ and $c_e$ are the start and end points of the current line segment being considered, and $c_i$ represents any intermediate point between them. The point $c_k$ is retained if its distance to the segment $(c_s, c_e)$ exceeds $\epsilon_{rdp}$, and the algorithm is recursively applied.

{\textbf{(b) Shift.}} The scarcity of long-tail data means that a specific rare event may appear only once in the training set. To mitigate overfitting to this single instance and increase the diversity of rare samples, we apply a spatial shift. This method adds a uniform random displacement to the entire trajectory, creating synthetic yet plausible variations of the same rare maneuver. This approach teaches the model to recognize the intrinsic pattern of the maneuver, decoupling it from its absolute spatial coordinates and enhancing its ability to generalize to similar unseen long-tail scenarios.
\begin{equation}
T'_{o} = \{c_1 + \Delta, c_2 + \Delta, \dots, c_{t_o} + \Delta\}
\end{equation}
where $\Delta = (\delta_x, \delta_y)$ is a displacement vector, and $\delta_x, \delta_y \sim \mathcal{U}(-\epsilon_{shift}, \epsilon_{shift})$. Here, $\epsilon_{shift}$ represents the dynamic shift magnitude provided by our strategy generator, which is strictly bounded by a predefined maximum threshold $\epsilon_{\max}$ (i.e., $\epsilon_{shift} \le \epsilon_{\max}$) to ensure the physical validity of the trajectory.

{\textbf{(c) Mask.}} Long-tail scenarios are often complex and may involve occlusions or sensor failures, resulting in incomplete observations. To ensure our model is robust in these high-stakes situations, we implement a masking augmentation. By randomly dropping points from the trajectory, we simulate partial data loss. This forces the model to infer the underlying intent of a rare maneuver even with imperfect information.
\begin{equation}
T'_{o} = 
\begin{cases} 
c_i, & \text{if } b_i = 1 \\ 
0, & \text{if } b_i = 0 
\end{cases}
\end{equation}
where the binary mask $b_i$ follows a Bernoulli distribution $\mathcal{B}(\rho)$, with $\rho$ representing the probability of retaining each point in the observed agent trajectory sequence.

{\textbf{(d) Subset.}} A rare event might unfold rapidly, or an agent might only enter the field of view midway through a critical maneuver. To train our model to recognize these developing long-tail events from limited temporal evidence, we employ subset selection. This method extracts a continuous sub-sequence from the trajectory, simulating scenarios of partial temporal observation. It enhances the model's ability to identify the onset of a rare behavior from short temporal cues, enabling earlier and more reliable predictions in dynamic, safety-critical situations.
\begin{equation}
T'_o = \{c_i, \dots, c_{i+\gamma*n}\}
\end{equation}
where $\gamma$ is the subset ratio, representing the proportion of the trajectory retained in the subset.

\begin{figure}[pos=t]
  \centering
\includegraphics[width=0.65\linewidth]{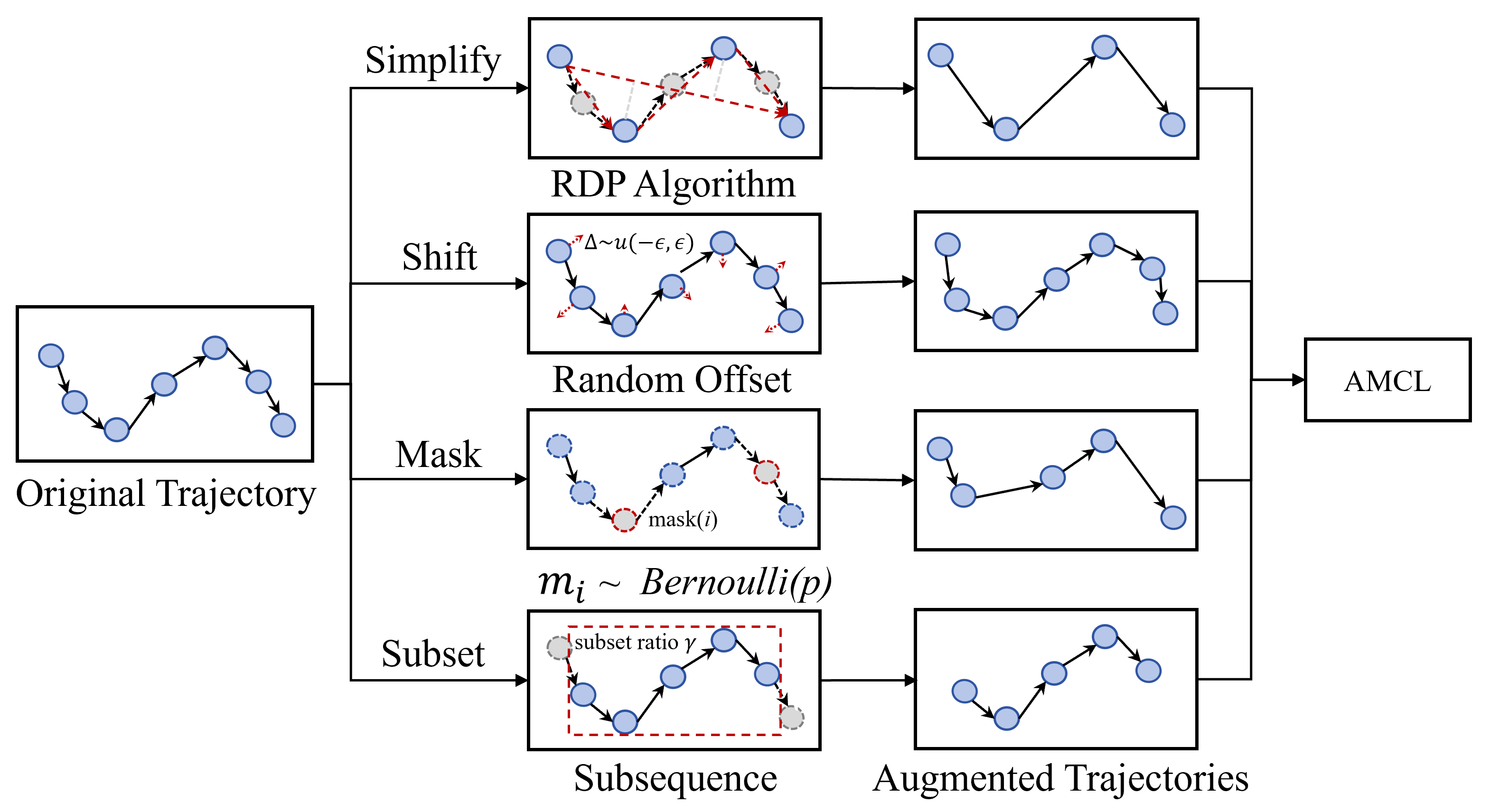} 
  \caption{Visualization of our Attribute-Guided Trajectory Augmentation strategies. Based on the identified long-tail attributes of a trajectory, AGTA applies a combination of targeted augmentations (Simplify, Shift, Mask, Subset) to create a diverse set of challenging positive samples for the subsequent contrastive learning stage.}
  \label{fig3} 
\end{figure}

\subsection{Scene Representation Learning}

\subsubsection{Context Encoding and Interaction}

The Motion and Context Encoder transforms raw observations into high-dimensional feature vectors. We employ a shared hierarchical architecture: historical agent states $H_{a}$ are processed through an embedding layer, a Transformer encoder for temporal dependencies, and a GRU to produce agent features $F_a = \{F_{\text{target}}, F_{\text{neighbors}}\}$. Simultaneously, the map vectors $M$ and their adjacency matrix $A_M$ are processed using a Graph Attention Network (GAT) to model the road topology, generating map features $F_m$.

To capture the inherent ambiguity of future scenarios, we introduce a {Scene Interactor} equipped with a latent variable mechanism to generate $K$ distinct interaction hypotheses. The target agent's feature $F_{\text{target}}$ is refined and projected to produce $K$ distinct modal queries $H_{\text{mode}} \in \mathbb{R}^{K \times D}$. These queries then selectively integrate information from the global environment through Multi-Head Attention ($\phi_{\text{MHA}}$), augmented with positional encoding $F_p$:
\begin{equation}
F_{\text{context}} = \phi_{\text{MHA}}\big(H_{\text{mode}}, [F_a, F_m] + F_p, [F_a, F_m] + F_p\big)
\end{equation}

This interaction process generates a rich, multimodal context feature set $F_{\text{context}} \in \mathbb{R}^{K \times D}$ for the target agent, representing $K$ plausible interactive intents ready for subsequent decoding.

\subsubsection{Attribute Disentanglement and Prediction}
We first aggregate the multimodal context features $F_{\text{context}}$ into a unified scene-level representation $F_{\text{scene}} \in \mathbb{R}^{D}$ using average pooling over the $K$ modes. This $F_{\text{scene}}$ serves as the foundation for subsequent attribute disentanglement. Inspired by advances in disentangled representation learning, our Attribute Feature Extractor separates the features associated with each long-tail attribute, moving beyond a simple, holistic scene understanding to a fine-grained, attribute-aware interpretation.
The extractor employs three independent self-attention mechanisms, each focusing on a specific attribute: Prediction Error, Collision Risk, and State Complexity. This enables each attention head to selectively emphasize the dimensions of $F_{\text{scene}}$ that are most relevant to its assigned attribute:
\begin{equation}
F_e = \phi_{\text{MHA}_e}(F_{\text{scene}}), \quad
F_r = \phi_{\text{MHA}_r}(F_{\text{scene}}), \quad
F_s = \phi_{\text{MHA}_s}(F_{\text{scene}}) 
\end{equation}

To explicitly define the semantic meaning of these latent spaces, each feature is passed through a dedicated lightweight MLP head to predict its corresponding ground-truth attribute value $(\hat{y}_e, \hat{y}_r, \hat{y}_s)$:
\begin{equation}
\hat{y}_e = \phi_{\text{MLP}_e}(F_e), \quad
\hat{y}_r = \phi_{\text{MLP}_r}(F_r), \quad
\hat{y}_s = \phi_{\text{MLP}_s}(F_s)
\end{equation}

These predictions are dynamically optimized via an auxiliary loss ($L_{{attr}}$) during training. This explicit supervision provides attribute-level guidance for each latent sub-space to align with its corresponding physical or model-level target. Such a structural constraint encourages the model to learn more disentangled and semantically meaningful representations, mitigating the feature entanglement commonly observed in unconstrained self-attention mechanisms. As a result, $F_e$, $F_r$, and $F_s$ are encouraged to encode different, yet not strictly independent, contextual information related to prediction difficulty, collision risk, and state complexity, respectively.


\subsection{Adaptive Contrastive Representation Learning}
In long-tail trajectory prediction, standard contrastive learning frameworks like MoCo \cite{he2020momentum-moco} face limitations. Their static momentum updates and reliance on random negative sampling can lead to suboptimal performance on rare patterns, as the learning signal is often dominated by easy negatives from the head of the distribution. To overcome these limitations, we propose Adaptive Momentum Contrastive Learning (AMCL). Our approach enhances representation learning through a continuous cosine momentum schedule for smoother training dynamics and a novel Similarity-Weighted Hard-Negative Mining strategy to intelligently focus on the most challenging samples. The operational details are outlined in Algorithm \ref{AMCL_Algorithm}. We replace fixed momentum schedules with a continuous cosine update mechanism. This allows the momentum coefficient $m$ to evolve smoothly throughout training, balancing feature plasticity and stability more effectively than a static approach. The momentum $m$ at training epoch $e$ is defined as:
\begin{equation}
m(e) = m_{f} - (m_{f} - m_{b}) \left( \frac{1 + \cos(\pi e / E_{\max})}{2} \right)
\end{equation}
where $m_{b}$ and $m_{f}$ are the initial and final momentum values, and $E_{\max}$ is the total training duration. This ensures a gradual and stable transition of the momentum encoder's parameters $\theta_k$:
\begin{equation}
\theta_k \leftarrow \theta_k - (1 - m(e)) \cdot (\theta_k - \theta_q)
\end{equation}

To refine the hard-negative mining process, we introduce a more nuanced mechanism. Recognizing that some hard negatives are more confusable than others, our approach assigns higher penalties to negatives that are more similar to the query, rather than treating them with equal importance. After identifying the top-$N_{neg}$ hard negatives $\{k_1^-, \dots, k_{N_{neg}}^-\}$ and their similarities $\{s_1^-, \dots, s_{N_{neg}}^-\}$ to the query $q$, we compute a weight $w_i$ for each negative using a softmax function:
\begin{equation}
w_i = \frac{\exp(s_i^- / \tau_w)}{\sum_{j=1}^{N_{neg}} \exp(s_j^- / \tau_w)}
\end{equation}
where $\tau_w$ is a temperature hyperparameter controlling the weight distribution. These weights are then incorporated into our final contrastive loss, which focuses the learning signal on the most challenging distinctions:
\begin{equation}
L_{{amcl}} = -\log \frac{\exp(s^+/\tau_a)}{\exp(s^+/\tau_a) + \sum_{i=1}^K w_i \cdot \exp(s_i^-/\tau_a)}
\end{equation}
where $s^+$ is the similarity of the positive pair and $\tau_a$ is the standard temperature parameter. This design forces the model to learn a more discriminative feature space for challenging long-tail patterns.

\begin{algorithm}[htbp]
\caption{Adaptive Momentum Contrastive Learning (AMCL)}
\label{AMCL_Algorithm}
\KwIn{Query encoder $\theta_q$, momentum encoder $\theta_k$, momentum params $m_{b}, m_{f}$, total duration $E_{\max}$, num hard negatives $N_{neg}$, temperatures $\tau_a, \tau_w$}
\KwOut{AMCL loss $L_{amcl}$}
\BlankLine
Initialize negative sample queue $Q$ \\
\For{$e \gets 1$ to $E_{\max}$}{
    Update momentum coefficient: $m \gets m_{f} - (m_{f} - m_{b}) \left( \frac{1 + \cos(\pi e / E_{\max})}{2} \right)$ \\
    Sample original sequence $x_q$ and positive $x_k \gets \texttt{Augment}(x_q)$ \\
    Encode query $q \gets \texttt{Encoder}_{\theta_q}(x_q)$ and positive $k^+ \gets \texttt{Encoder}_{\theta_k}(x_k)$ \\
    Update momentum encoder: $\theta_k \gets m \cdot \theta_k + (1 - m) \cdot \theta_q$ \\
    Compute positive similarity: $s^+ \gets \texttt{Similarity}(q, k^+)$ \\
    Select top-$N_{neg}$ hard negatives $\{k_i^-\}_{i=1}^{N_{neg}}$ and similarities $\{s_i^-\}_{i=1}^{N_{neg}}$ from $Q$ \\
    Compute weights for hard negatives: $w_i \gets \frac{\exp(s_i^- / \tau_w)}{\sum_{j=1}^{N_{neg}} \exp(s_j^- / \tau_w)}$ for $i=1...N_{neg}$ \\
    Compute weighted denominator: $D_{neg} \gets \sum_{i=1}^{N_{neg}} w_i \cdot \exp(s_i^- / \tau_a)$ \\
    Compute AMCL loss: $L_{amcl} \gets -\log \frac{\exp(s^+ / \tau_a)}{\exp(s^+ / \tau_a) + D_{neg}}$ \\
    Optimize query encoder: $\texttt{Optimize}_{\theta_q}(L_{amcl})$ and append $k^+$ to $Q$ \\
    \If{$|Q| > Q_{\max}$}{
        Remove oldest sample: $Q \leftarrow Q \setminus q_{\text{oldest}}$
    }
}
\Return $L_{amcl}$
\end{algorithm}

\subsection{Progressive Pseudo-Label Supervision}
To identify long-tail trajectory patterns within the feature space, many studies employ clustering to generate feature labels. However, these should be regarded as pseudo-labels, as their validity is not static. A critical oversight in many approaches is treating them as fixed ground truth throughout training. This approach fails to account for the non-stationary nature of the feature space, where representations for rare patterns evolve significantly and only become separable after extensive training. To address this, we propose the Evolving Feature Clustering (EFC) strategy, which synchronizes the pseudo-labels with the model's evolving feature manifold.

The EFC strategy operates by first accumulating the encoded feature representations into a comprehensive memory bank $\mathcal{B} = \{f_i\}_{i=1}^N$. Instead of a one-time clustering, EFC is periodically activated at predefined intervals. At each activation epoch $e$, the K-means algorithm \cite{hartigan1979algorithm-kmeans} is applied to the entire memory bank $\mathcal{B}$ to find $C$ new cluster centroids $\mathcal{P}^{(e)} = \{\mu_j\}_{j=1}^C$ by minimizing the intra-cluster variance:
\begin{equation}
\min_{\mathcal{P}^{(e)}} \sum_{j=1}^C \sum_{f_i \in \mathcal{C}_j} \| f_i - \mu_j \|^2
\label{eq:kmeans_objective}
\end{equation}
where $\mathcal{C}_j$ denotes the $j$-th cluster. Once the optimal centroids are found, each feature $f_i$ in the memory bank is assigned a new pseudo-label ${PL}_i^{(e)}$ based on its nearest centroid:
\begin{equation}
{PL}_i^{(e)} = \text{argmin}_{j \in \{1, ..., C\}} \| f_i - \mu_j \|^2
\label{eq:kmeans_assignment}
\end{equation}

These dynamically updated pseudo-labels ${PL}^{(e)} = \{{PL}_i^{(e)}\}_{i=1}^N$ are then used as the supervisory signal for subsequent Decoupled Contrastive Learning. To ensure the stability and quality of the generated pseudo-labels, the EFC process is introduced progressively. In particular, explicit clustering is bypassed during an initial warm-up phase. During this early training stage, feature representations are optimized by the unsupervised AMCL module. This approach allows the feature manifold to reach a preliminary level of discriminability before hard cluster assignments are enforced, mitigating the risk of generating fluctuating or noisy pseudo-labels from a premature feature space. By synchronizing the pseudo-labeling with the model's learning state, EFC provides increasingly accurate supervision.

\subsection{Focused Decoupled Contrastive Learning}
With the dynamic pseudo-labels from our EFC strategy, we need a supervised contrastive loss that can handle the severe class imbalance inherent in long-tail data. While Decoupled Contrastive Learning \cite{xuan2024decoupled} effectively addresses inter-class imbalance by re-weighting based on class size, it treats all intra-class positive pairs with equal importance. This can be suboptimal for long-tail classes, which often exhibit high intra-class variance. To address this, we propose Focused Decoupled Contrastive Learning (FDCL), an enhanced method that dynamically focuses on hard-positive samples. FDCL introduces a focusing weight $w_f$, inspired by Focal Loss, which modulates the attractive force between positive pairs based on their similarity. This is combined with Decoupled Contrastive Learning's original class-aware weight $w_r$, which is defined as:
\begin{equation}
w_r(q_t) =
\begin{cases}
\alpha(|\mathcal{P}_i| + 1), & \text{if } q_t = q_i^+ \\
(1 - \alpha)(|\mathcal{P}_i| + 1)/{|\mathcal{P}_i|}, & \text{if } q_t \in \mathcal{P}_i
\end{cases}
\end{equation}
where $\mathcal{P}_i$ is the set of other positive features sharing the same pseudo-label as the query $q_i$, and $q_i^+$ is the differently augmented view of the query. We formulate the final combined weight $w_{fdcl}$ for any target positive sample $q_t$ as:
\begin{equation}
w_{fdcl}(q_t) = w_r(q_t) \cdot w_f(q_t) = w_r(q_t) \cdot (1 - \langle q_i, q_t \rangle)^\eta
\end{equation}
where $\langle \cdot, \cdot \rangle$ denotes cosine similarity, and $\eta$ is a focusing hyperparameter. By penalizing easy positives (high similarity) and amplifying hard positives (low similarity), the final FDCL loss is computed as:
\begin{equation}
L_{fdcl} = \frac{-1}{|\mathcal{P}_i| + 1} \sum_{q_t \in \{q_i^+\} \cup \mathcal{P}_i} \log \frac{\exp(w_{fdcl}(q_t) \cdot \langle q_i, q_t \rangle / \tau_f)}{\sum_{q_m \in \{q_t\} \cup \mathcal{N}_i} \exp(\langle q_i, q_m \rangle / \tau_f)}
\end{equation}
where $\mathcal{N}_i$ is the set of negative samples from other pseudo-label classes, and $\tau_f$ is the temperature hyperparameter. By mathematically encouraging the model to focus more on dissimilar positives, FDCL encourages the formation of more compact and well-separated clusters, which is particularly beneficial for the diverse and fragmented patterns found in tail-end classes. The detailed procedure is outlined in Algorithm \ref{fdcl}.

\begin{algorithm}[htbp]
\caption{Focused Decoupled Contrastive Learning (FDCL)}
\label{fdcl}
\KwIn{Query $q_i$, augmented positive $q_i^+$, positive set $\mathcal{P}_i$, negative set $\mathcal{N}_i$, temperature $\tau_f$, parameters $\alpha, \eta$}
\KwOut{FDCL Loss $L_{fdcl}$}
\BlankLine
Initialize batch loss: $L_{fdcl} \gets 0$ \\
\For{each query feature $q_i$ in batch}{
    L2-Normalize all features: $q_i, q_i^+, \mathcal{P}_i, \mathcal{N}_i$ \\
    Construct full positive set: $\mathcal{S}_i \gets \{q_i^+\} \cup \mathcal{P}_i$ \\
    Initialize sample loss: $L_i \gets 0$ \\
    \For{each positive feature $q_t \in \mathcal{S}_i$}{
        Compute class-aware weight: $w_r \gets \alpha (|\mathcal{P}_i| + 1)$ \textbf{if} $q_t = q_i^+$ \textbf{else} $(1 - \alpha)(|\mathcal{P}_i| + 1)/{|\mathcal{P}_i|}$ \\
        Compute final focusing weight: $w_{fdcl} \gets w_r \cdot (1 - \langle q_i, q_t \rangle)^\eta$ \\
        Compute weighted numerator: $N_{i,t} \gets \exp(w_{fdcl} \cdot \langle q_i, q_t \rangle / \tau_f)$ \\
        Compute decoupled denominator: $D_{i,t} \gets \sum_{q_m \in \{q_t\} \cup \mathcal{N}_i} \exp(\langle q_i, q_m \rangle / \tau_f)$ \\
        Update sample loss: $L_i \gets L_i - \log(N_{i,t} / D_{i,t})$ \\
    }
    Accumulate batch loss: $L_{fdcl} \gets L_{fdcl} + \frac{L_i}{|\mathcal{P}_i| + 1}$ \\
}
\Return{ $L_{fdcl}$}
\end{algorithm}

\subsection{Attribute-aware Trajectory Generator}
The final stage of our framework is the Attribute-aware Trajectory Generator, which is responsible for synthesizing multiple plausible future paths based on the rich, fused representations from the upstream modules. This generator is explicitly designed to leverage the disentangled attribute features, enabling it to produce more informed and context-aware predictions, especially in complex long-tail scenarios. Central to this module is an attribute-gating mechanism that dynamically weights the influence of different long-tail attributes. First, a set of dynamic gating weights ${g}$ is generated from the unified scene representation ${F}_{\text{scene}}$:
\begin{equation}
g = [g_e, g_r, g_s]^\top = \sigma(\phi_g(F_{\text{scene}}))
\end{equation}
where $\phi_g$ is an MLP and $\sigma$ is the sigmoid function. These gates then modulate the contribution of each disentangled attribute feature ($F_e, F_r, F_s$) to form a weighted, attribute-aware feature $F_{\text{gated}}$:
\begin{equation}
F_{\text{gated}} = g_e \cdot F_e + g_r \cdot F_r + g_s \cdot F_s
\end{equation}

Subsequently, this attribute-aware feature $F_{\text{gated}}$ is fused with the multimodal context features $F_{\text{context}}$ produced by the Scene Interactor via element-wise addition, yielding the enhanced representation $F_{\text{fused}}$:
\begin{equation}
F_{\text{fused}} = F_{\text{context}} + F_{\text{gated}}
\end{equation}

The generator then processes this final fused representation $F_{\text{fused}}$ to produce the trajectories. A GRU acts as the decoder, taking $F_{\text{fused}}$ as its initial hidden state $h_0$ and sequentially generating the parameters for the future trajectory distribution at each time step $t \in (t_o, t_o + t_p]$:
\begin{equation}
h_t = \phi_\text{GRU}(h_{t-1}, z_t)
\end{equation}
where $z_t$ is a learnable input query for each time step. The output of the GRU at each step is then passed to a MLP layer $\phi_\text{MLP}$ to predict the parameters of a Laplace Mixture Density Network (MDN). For each of the $K$ modes, the network outputs the spatial coordinate estimate ${c}_t^k$, scale parameter ${s}_t^k$, and the probability $\pi^k$ for the trajectory mode:
\begin{equation}
({c}^k_t, {s}^k_t), {\pi}^k = \phi_\text{MLP}(h_t, h_0)
\end{equation}

\subsection{Loss Function}
The training of our model is guided by a comprehensive, multi-component objective function designed to address prediction accuracy, attribute disentanglement, and robust feature representation simultaneously.

The primary prediction objective \( L_{{task}} \) is a composite loss that handles the multimodal nature of the task. It integrates three key terms: a Laplace negative log-likelihood \( L_{{reg}} \) for the regression of trajectory coordinates, a cross-entropy loss \( L_{{cls}} \) for classifying the most likely prediction mode, and a target loss \( L_{{target}} \) defined by the minimum Average Displacement Error (minADE) over the $K$ modes. These are combined as:
\begin{equation}
L_{{task}} = L_{{target}} + L_{{reg}} + L_{{cls}}
\end{equation}

To ensure our model effectively disentangles the specified long-tail attributes, we introduce an auxiliary supervision signal \( L_{{attr}} \). This loss penalizes the discrepancy between the predicted attributes and their corresponding ground-truth. We employ a Mean Squared Error (MSE) loss for this purpose:
\begin{equation}
L_{{attr}} = \| [\hat{y}_e, \hat{y}_r, \hat{y}_s] - [y_e, y_r, y_s] \|_2^2
\label{eq:aux_loss}
\end{equation}

Additionally, to enforce the learning of discriminative features, we incorporate the two contrastive losses central to our framework: the adaptive momentum contrastive loss \( L_{{amcl}} \) and the focused decoupled contrastive loss \( L_{fdcl} \). The overall training objective $L$ is a weighted sum of all these components:
\begin{equation}
L = L_{{task}} + \lambda_1 L_{{attr}} + \lambda_2 L_{{amcl}} + \lambda_3 L_{{fdcl}}
\end{equation}
where \( \lambda_1 \), \( \lambda_2 \), and \( \lambda_3 \) are hyperparameters that balance the contribution of each loss term.

\section{Experiments} \label{Experiments}
\subsection{Experimental Setups}
\subsubsection{Datasets} 
To validate the effectiveness and robustness of our method, we conduct experiments on two widely recognized benchmarks: nuScenes \cite{caesar2020nuscenes} for autonomous driving and ETH/UCY \cite{pellegrini2009you, leal2014learning} for pedestrian trajectory prediction. These datasets provide a diverse range of real-world traffic scenarios. 

\textbf{nuScenes}:  This comprehensive autonomous driving benchmark features 1000 distinct scenes, capturing complex real-world driving conditions supported by detailed HD maps. The dataset provides 2 s of historical trajectory data and 6 s of future trajectory data for target agent, making it suitable for evaluating various prediction horizons.
 
\textbf{ETH/UCY}: Focusing on pedestrian dynamics, this benchmark serves as a standard for crowd behavior analysis. We specifically include this dataset to validate the cross-agent robustness of our model, recognizing that safe autonomous driving necessitates accurate prediction of both vehicle and pedestrian trajectories in complex, interactive environments. It aggregates five unique environments: ETH and HOTEL (from the ETH dataset), along with UNIV, ZARA1, and ZARA2 (from the UCY dataset). The data is recorded at 2.5 Hz. In our experiments, we observe 8 timesteps (equivalent to 3.2 s) to forecast the subsequent 12 timesteps (equivalent to 4.8 s).

\subsubsection{Long-tail Evaluation Subsets} 
To validate our model's performance on long-tail data, we construct dedicated long-tail evaluation subsets based on the three distinct attributes. This multi-dimensional approach to subset creation allows for a more fine-grained assessment of our model's ability to handle specific long-tail challenges, which differs significantly from previous studies that often rely on single-dimensional or heuristic-based definitions of long-tail scenarios. The dataset is systematically divided into distinct subsets for each attribute:
\begin{itemize}
    \setlength{\leftskip}{0.5em}
     \item \textbf{Prediction Error}: We sort all samples by their $y_e$ values in descending order. The long-tail subset consists of the Top 1\%-5\% of samples exhibiting the highest prediction errors. This represents scenarios that are intrinsically difficult to predict by a baseline model.
    
    \item \textbf{Collision Risk}: We sort all samples by their $y_r$ values in descending order. The long-tail subset comprises the Top 1\%-5\% of samples with the highest InvTTC values, indicating the most safety-critical scenarios.
    
    \item \textbf{State Complexity}: For enhanced interpretability, we discretize the continuous $y_s$ metric into several distinct behavior categories: Abrupt Acceleration, Abrupt Deceleration, Abrupt Lane Changes, and High-Curvature Turns. Samples not falling into these categories are considered Normal behaviors.
\end{itemize}

\subsubsection{Evaluation Metrics}

To assess the predictive fidelity and robustness of our framework, we adopt standard trajectory prediction metrics following the evaluation protocol of each benchmark. 
For ETH/UCY and the long-tail comparison on nuScenes, we report minADE and minFDE for consistency with existing methods. 
For the full-sample comparison on nuScenes, we follow the official benchmark protocol and prior literature, and report $\text{minADE}_{5/10}$, $\text{minFDE}_{1/5/10}$, and $\text{MR}_5$. 
Below, we provide the detailed definitions and formulas for these metrics.

\begin{itemize}
    \item \textbf{Minimum Average Displacement Error (minADE):} 
    Given $K$ predicted trajectories $\{\hat{T}^{(k)}\}_{k=1}^{K}$ and the ground-truth trajectory $T = \{ c_t \mid t \in (t_o, t_o+t_p] \}$, minADE is defined as the minimum average displacement error among all predicted modes:
    \begin{equation}
    \text{minADE}_K = \min_{k=1,\dots,K} \left[ \frac{1}{t_p} \sum_{t=1}^{t_p} \| \hat{c}_t^{(k)} - c_t \|_2 \right]
    \end{equation}

    \item \textbf{Minimum Final Displacement Error (minFDE):} 
    minFDE is defined as the minimum final displacement error among all predicted modes:
    \begin{equation}
    \text{minFDE}_K = \min_{k=1,\dots,K} \| \hat{c}_{t_p}^{(k)} - c_{t_p} \|_2 
    \end{equation}

    \item \textbf{Miss Rate (MR):} 
    MR measures the fraction of samples for which none of the predicted trajectories falls within a threshold $\delta$ (typically set to 2 m) of the ground-truth endpoint:
    \begin{equation}
    \text{MR}_K = \frac{1}{N} \sum_{i=1}^{N}
    I\!\left( \min_{k=1,\dots,K} \| \hat{c}_{t_p}^{i,(k)} - c_{t_p}^{i} \|_2 > \delta \right)
    \end{equation}
    where $I(\cdot)$ is the indicator function.
\end{itemize}

\subsubsection{Implementation Details}

All experiments are implemented in PyTorch and conducted on a single NVIDIA RTX 3090 GPU. The encoder embedding dimension is set to 32, with three Transformer encoder layers, two GAT layers, and four attention heads in the scene interaction module. The number of predicted trajectories is set to $K=25$. In AMCL, the momentum coefficients $m_b$ and $m_f$ are set to 0.95 and 0.999, respectively. The model is trained using the Adam optimizer with a learning rate of 0.0005 and a batch size of 32. The loss weights $\lambda_1$, $\lambda_2$, and $\lambda_3$ are set to 1, 1, and 0.1, respectively. The total training process lasts for 120 epochs, including a 10-epoch warm-up stage before clustering is enabled. K-means clustering is performed every 5 epochs to update pseudo-labels based on the evolving feature distribution, and the number of clusters is set to 5.

\begin{table*}[htbp]
\centering
\caption{
 Prediction results (minADE/minFDE) on nuScenes (2 s prediction) and ETH/UCY (4.8 s prediction) datasets. Samples are stratified based on their prediction error (FDE) to evaluate robustness, with the Top 1\%-5\% representing the highest error instances. \textbf{Bold} and \uline{underlined} text indicate the best and second-best results, respectively. Cases marked with ('-') indicate missing values. Improvement row shows the percentage improvement of our model over the \uline{second-best} result.
 }
\resizebox{1\linewidth}{!}{
\begin{tabular}{l|cccccc|c} 
\toprule
 Model                   & Top 1\%                   & Top 2\%                   & Top 3\%                   & Top 4\%                   & Top 5\%                   & Rest                      & All                         \\ 
\midrule
\multicolumn{8}{c}{\textbf{nuScenes Dataset}}\\
\midrule
 Traj++ EWTA \cite{makansi2021on-exposing}            & 1.73/4.43                & 1.36/3.54                & 1.17/3.03                & 1.04/2.68                & 0.95/2.41                & 0.16/0.26                 & 0.22/0.39                   \\ 

 Traj++ EWTA+contrastive \cite{makansi2021on-exposing} & 1.28/2.85                 & 0.97/2.15                 & 0.83/1.83                 & 0.76/1.64                 & 0.70/1.48                 & \uline{0.15}/0.24                 & {0.18}/0.30           \\ 

 FEND \cite{wang2023fend}                   & {1.21}/{2.50} & {0.92}/{1.88} & {0.79}/{1.61} & {0.72}/{1.43} & {0.66}/{1.31} & \textbf{0.14}/\uline{0.20} & \uline{0.17}/\uline{0.26}  \\ 

 TrACT \cite{zhang2024tract}                  & 1.23/2.65                & 0.98/2.11                & 0.85/1.82                & 0.78/1.64                & 0.72/1.49                & -                         & 0.19/0.31                  \\ 
CSD \cite{ganeshaaraj2025enhancing}                  & \uline{1.06}/\uline{2.22}                & \uline{0.82}/\uline{1.69}                & \uline{0.70}/\uline{1.44}                & \uline{0.64}/\uline{1.30}                & \uline{0.59}/\uline{1.18}                & -                         & {0.18}/0.30                  \\

SAIL (Ours) & \textbf{1.02}/\textbf{1.58} & \textbf{0.81}/\textbf{1.30} & \textbf{0.65}/\textbf{1.12} & \textbf{0.60}/\textbf{1.05} & \textbf{0.56}/\textbf{0.98} & 0.16/{\textbf{0.19}} & \textbf{0.17}/\textbf{0.23} \\
\midrule
 \textit{Improvement} & \dualimpr{+3.8}{+28.8} & \dualimpr{+1.2}{+23.1} & \dualimpr{+7.1}{+22.2} & \dualimpr{+6.3}{+19.2} & \dualimpr{+5.1}{+16.9} & \dualimpr{-14.3}{+5.0} & \dualimpr{0.0}{+11.5} \\
\midrule
\multicolumn{8}{c}{\textbf{ETH/UCY Dataset}}  \\
\midrule
 Traj++ EWTA \cite{makansi2021on-exposing}            & 0.98/2.54                & 0.79/2.07                & 0.71/1.81                & 0.65/1.63                & 0.60/1.50                &\textbf{0.14}/\uline{0.26}                 & 0.17/0.32                   \\ 
          Traj++ EWTA+resample \cite{shen2016relay} & 0.90/2.17                & 0.77/1.90                & 0.73/1.78                & 0.66/1.60                & 0.64/1.52                & 0.20/0.41                 & 0.23/0.47                   \\
          Traj++ EWTA+reweighting \cite{cui2019class} & 0.97/2.47 & 0.78/2.03 & 0.68/1.73 & 0.62/1.55 & 0.56/1.40 & 0.15/0.26 & 0.18/0.32 \\
         
          Traj++ EWTA+contrastive \cite{makansi2021on-exposing} & 0.92/2.33 & 0.74/1.91 & 0.67/1.71 & 0.60/1.48 & 0.55/1.32 & 0.15/0.27 & 0.17/0.32 \\
          LDAM \cite{cao2019learning}                 & 0.92/2.35                & 0.76/1.96                & 0.68/1.71                & 0.62/1.53                & 0.57/1.37                & 0.15/0.27                 & 0.17/0.33                   \\ 
          FEND \cite{wang2023fend}                   & 0.84/2.13                & 0.68/1.68                & \uline{0.61}/\uline{1.46}                & 0.56/\uline{1.30}                & \uline{0.52}/1.19                & \uline{0.15}/0.27                 & {0.17}/0.32                   \\ 
         TrACT \cite{zhang2024tract}                  & \uline{0.80}/\uline{2.00} & \textbf{0.65}/\uline{1.63} & \uline{0.61}/\uline{1.46} & \uline{0.56}/1.31 & \uline{0.52}/\uline{1.18} & -                 &\uline{0.17}/\uline{0.32}   \\ 
 SAIL (Ours) & \textbf{0.73}/\textbf{1.69} & \uline{0.66}/\textbf{1.58} & \textbf{0.59}/\textbf{1.38} & \textbf{0.55}/\textbf{1.27} & \textbf{0.51}/\textbf{1.17} & 0.17/\textbf{0.25} & \textbf{0.17}/\textbf{0.27} \\
 \midrule
\textit{Improvement} & \dualimpr{+8.8}{+15.5} & \dualimpr{-1.5}{+3.1} & \dualimpr{+3.3}{+5.5} & \dualimpr{+1.8}{+2.3} & \dualimpr{+1.9}{+0.8} & \dualimpr{-13.3}{+3.8} & \dualimpr{0.0}{+15.6} \\
\bottomrule
\end{tabular}
}
\label{table_2}%
\end{table*}

\subsection{Quantitative Analysis} 
\subsubsection{Quantitative Results on Prediction Error}

As presented in Table \ref{table_2}, we compare SAIL with several representative baseline methods specifically designed for long-tail trajectory prediction. The results show that SAIL achieves clear advantages on the most challenging subsets and establishes new state-of-the-art performance on severe long-tail cases. This is particularly evident on the nuScenes dataset, where for the hardest 1\% of cases, SAIL achieves a minFDE of 1.58 m. This constitutes a remarkable 28.8\% reduction in error compared to the next-best baseline CSD. Similarly, on the ETH/UCY dataset, our model reduces the minFDE by 15.5\% over the strongest competitor in this category. These substantial gains on the most extreme and unpredictable trajectories underscore the effectiveness of SAIL's adaptive learning mechanisms in capturing sparse, high-information patterns. At the same time, the results also reveal a characteristic trade-off. While SAIL shows the largest gains on the severe long-tail subsets (Top 1\%--5\%), its improvements on normal scenarios are less pronounced, and minADE on the ``Rest'' subset may show a slight decrease. This behavior is consistent with the design objective of the framework, which explicitly allocates more representational and optimization focus to rare, difficult, and high-risk cases. As a result, the model places less emphasis on further optimizing easy majority samples that are already well represented in the data distribution. Nevertheless, SAIL maintains strong overall performance, achieving the best overall minFDE of 0.23 m on nuScenes and 0.27 m on ETH/UCY. These results suggest that SAIL improves robustness on the most difficult cases without sacrificing competitive performance on the full dataset.

\begin{table*}[htbp]
\centering
\caption{Prediction results (minADE/minFDE) at different prediction horizons on the nuScenes dataset, based on Collision Risk. We compare our model with the baseline Q-EANet~\cite{chen2024q-qeanet}. The Top 1\%-5\% intervals denote the most critical risk categories. \textbf{Bold} marks the best metric, while light gray background highlights cases where {both} metrics are simultaneously the best.}
\resizebox{1\linewidth}{!}{
\begin{tabular}{c|c|cccccc|c}
\toprule
 Horizon (s) & Model   & Top 1\%                  & Top 2\%                  & Top 3\%                  & Top 4\%                  & Top 5\%                  & Rest                  & All                   \\
\midrule
\multirow{2}{*}{1} 
  & Q-EANet & \textbf{0.18}/{0.22} & {0.22}/{0.25} & \textbf{0.17}/0.21       & \textbf{0.15}/{0.19} & \textbf{0.13}/{0.17} & \cellcolor{gray!20}\textbf{0.08}/\textbf{0.10} & \textbf{0.13}/0.12    \\
  & SAIL    & 0.19/\textbf{0.21}       & \cellcolor{gray!20}\textbf{0.20}/\textbf{0.23} & 0.18/\textbf{0.20}       & 0.16/\textbf{0.18}       & 0.14/\textbf{0.16}       & 0.13/0.11            & 0.14/\textbf{0.12}    \\
\midrule
\multirow{2}{*}{2} 
  & Q-EANet & 0.35/0.45                & 0.45/0.52                & 0.37/0.48                & 0.35/0.43                & 0.30/0.39                & 0.18/0.20            & 0.22/0.23             \\
  & SAIL    & \cellcolor{gray!20}\textbf{0.32}/\textbf{0.40} & \cellcolor{gray!20}\textbf{0.42}/\textbf{0.48} & \cellcolor{gray!20}\textbf{0.36}/\textbf{0.45} & \cellcolor{gray!20}\textbf{0.33}/\textbf{0.40} & \cellcolor{gray!20}\textbf{0.28}/\textbf{0.37} & \cellcolor{gray!20}\textbf{0.16}/\textbf{0.19} & \cellcolor{gray!20}\textbf{0.18}/\textbf{0.23} \\
\midrule
\multirow{2}{*}{3} 
  & Q-EANet & 0.55/0.75                & 0.67/0.82                & 0.58/0.78                & 0.56/0.71                & 0.49/0.65                & \textbf{0.29}/0.33   & 0.34/0.38             \\
  & SAIL    & \cellcolor{gray!20}\textbf{0.50}/\textbf{0.70} & \cellcolor{gray!20}\textbf{0.63}/\textbf{0.78} & \cellcolor{gray!20}\textbf{0.57}/\textbf{0.73} & \cellcolor{gray!20}\textbf{0.53}/\textbf{0.68} & \cellcolor{gray!20}\textbf{0.47}/\textbf{0.63} & 0.30/\textbf{0.33}   & \cellcolor{gray!20}\textbf{0.32}/\textbf{0.34} \\
\midrule
\multirow{2}{*}{4} 
  & Q-EANet & 0.75/1.05                & 0.88/1.15                & 0.80/1.09                & 0.73/0.98                & \textbf{0.66}/0.90       & 0.33/0.46            & 0.46/0.57             \\
  & SAIL    & \cellcolor{gray!20}\textbf{0.70}/\textbf{0.98} & \cellcolor{gray!20}\textbf{0.81}/\textbf{1.10} & \cellcolor{gray!20}\textbf{0.78}/\textbf{1.05} & \cellcolor{gray!20}\textbf{0.71}/\textbf{0.95} & 0.68/\textbf{0.88}       & \cellcolor{gray!20}\textbf{0.30}/\textbf{0.41} & \cellcolor{gray!20}\textbf{0.42}/\textbf{0.48} \\
\midrule
\multirow{2}{*}{5} 
  & Q-EANet & 0.95/1.35                & \textbf{1.05}/1.50       & 1.03/1.42                & 0.95/1.28                    & \textbf{0.83}/1.18       & 0.52/0.62            & 0.59/0.78             \\
  & SAIL    & \cellcolor{gray!20}\textbf{0.88}/\textbf{1.25} & 1.08/\textbf{1.45}       & \cellcolor{gray!20}\textbf{1.00}/\textbf{1.38} & \cellcolor{gray!20}\textbf{0.90}/\textbf{1.25} & 0.88/\textbf{1.15}       & \cellcolor{gray!20}\textbf{0.48}/\textbf{0.55} & \cellcolor{gray!20}\textbf{0.54}/\textbf{0.65} \\
\midrule
\multirow{2}{*}{6} 
  & Q-EANet & 1.24/1.65                & \textbf{1.25}/1.80       & 1.21/1.72                & 1.14/1.58                & 1.07/1.45                & 0.65/0.86            & 0.70/0.99             \\
  & SAIL    & \cellcolor{gray!20}\textbf{1.14}/\textbf{1.50} & 1.28/\textbf{1.70}       & \cellcolor{gray!20}\textbf{1.21}/\textbf{1.65} & \cellcolor{gray!20}\textbf{1.13}/\textbf{1.52} & \cellcolor{gray!20}\textbf{1.05}/\textbf{1.40} & \cellcolor{gray!20}\textbf{0.62}/\textbf{0.79} & \cellcolor{gray!20}\textbf{0.69}/\textbf{0.88} \\
\bottomrule
\end{tabular}
}
\label{table_3}
\end{table*}

\begin{figure}[pos=t]
  \centering
  \includegraphics[width=1\linewidth]{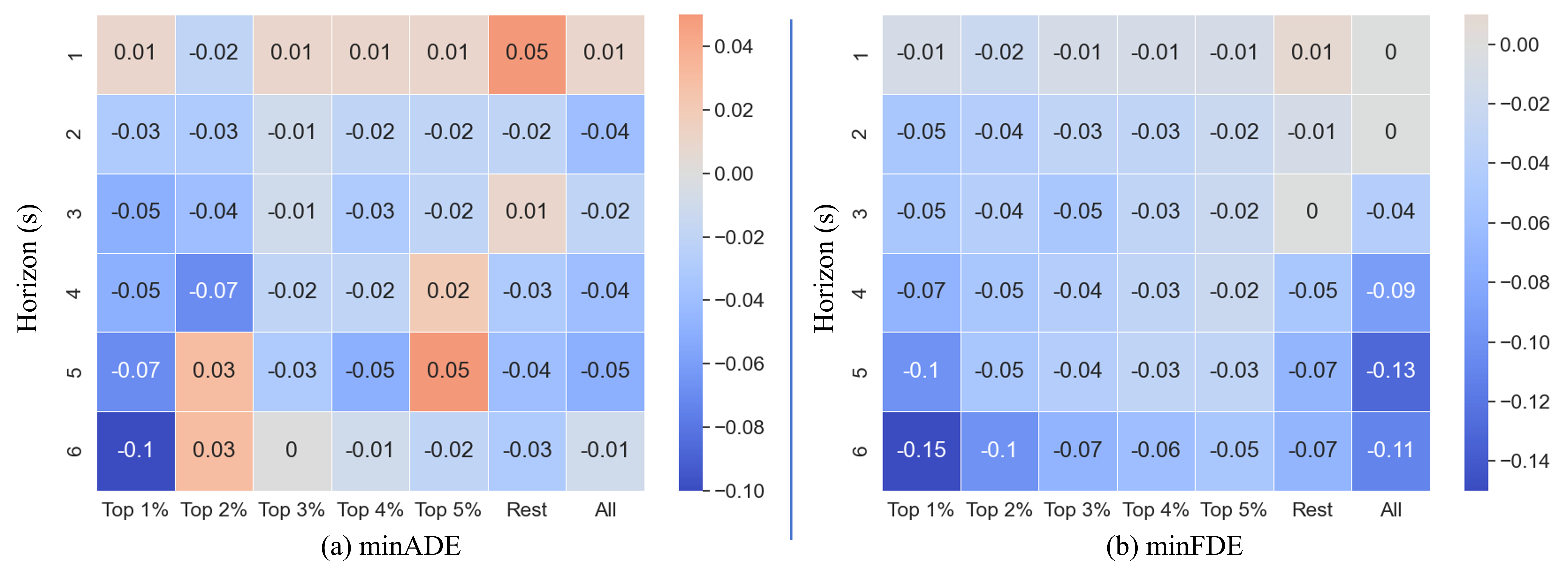}
   \caption{Heatmaps illustrating the performance improvements of the SAIL model relative to Q-EANet on the nuScenes dataset, categorized by collision risk levels and prediction horizons. Negative values indicate superior performance by SAIL. (a) Differences in minADE. (b) Differences in minFDE.}
   \label{fig4}
\end{figure}

\subsubsection{Quantitative Results on Collision Risk}
To further validate our model's effectiveness from a safety-critical perspective, we conduct a quantitative analysis based on a collision risk metric based on InverTTC. As shown in Table \ref{table_3}, the trajectories are categorized into risk levels, with the Top 1-5\% representing the most dangerous samples. Within SAIL's predictions, we observe that higher collision risk generally correlates with greater prediction error, as these scenarios often involve complex interactions and non-linear behaviors that are challenging to forecast. However, this relationship is not perfectly linear. For instance, at the 1\,s and 2\,s horizons, the prediction error for the Top 2\% risk category is slightly higher than for the Top 1\%. This suggests that while the Top 1\% scenarios are flagged as the most dangerous, they may include some predictably hazardous behaviors such as consistent close-proximity following in traffic, whereas slightly lower risk tiers could encompass more erratic maneuvers like abrupt speed variations that pose unique modeling difficulties.

Building on this internal analysis, a comparative evaluation against the Q-EANet \cite{chen2024q-qeanet} baseline reveals how predictive performance diverges as the forecast horizon increases. At the short 1 s horizon, where future motion is highly constrained and uncertainty is minimal, both models exhibit competitive performance with only marginal differences. This suggests the task at this range is not challenging enough to distinguish the capabilities of advanced models clearly. However, a clear and consistent trend emerges from the 2 s horizon onwards: SAIL begins to establish a distinct advantage, particularly in the safety-critical minFDE metric across the high-risk categories. This performance gap widens substantially as the horizon extends, a trend visually confirmed by the heatmaps in Figure \ref{fig4}. The increasingly negative values in the heatmap at longer horizons underscore our model's superior ability to handle compounding uncertainty. The advantage is most pronounced in high-risk, long-range scenarios; for example, at the 6 s horizon for Top 1\% trajectories, SAIL’s minFDE of 1.50 represents a substantial 9.1\% error reduction over the baseline. These results demonstrate that SAIL not only excels in standard short-term prediction but also maintains its robustness in longer, more uncertain forecasting horizons. This capability is essential for real-world autonomous systems, where reliable long-term planning is paramount for ensuring safety.

\subsubsection{Quantitative Results on State Complexity}

To further dissect our model's ability to handle diverse long-tail scenarios, we evaluate its performance on trajectories categorized by the complexity of their motion state. As detailed in Table \ref{table_4}, these states represent common yet challenging real-world driving behaviors. The results in Table \ref{table_4} demonstrate SAIL's robust performance across this spectrum of complexity. Our model consistently achieves the best overall prediction accuracy, leading in the ``All'' category across all prediction horizons. More importantly, SAIL shows a pronounced advantage in the most challenging motion states. For Abrupt Lane Changes and High-Curvature Turns scenarios, our model significantly outperforms the baseline, especially in the safety-critical minFDE metric. For instance, at the 6 s horizon, SAIL achieves a substantial 14.0\% reduction in minFDE for Abrupt Lane Changes. This highlights that SAIL is specifically optimized to excel at modeling the high-complexity, high-uncertainty dynamics of the most challenging long-tail events, making it a more reliable choice for navigating the diverse complexities of real-world traffic.

\begin{table*}[htbp]
\centering
\caption{Prediction results (minADE/minFDE) at different prediction horizons on the nuScenes dataset, based on vehicle state. We compare our model with the baseline Q-EANet~\cite{chen2024q-qeanet}. \textbf{Bold} indicates the best result for each metric. \cellcolor{gray!20}Light gray background highlights cases where {both} metrics are simultaneously the best.}

\resizebox{0.95\linewidth}{!}{
\begin{tabular}{c|c|ccccc|c}
\toprule
Horizon (s) & Model   & Abrupt Acceleration       & Abrupt Deceleration       & Abrupt Lane Changes        & High-Curvature Turns               & Normal                   & All                      \\
\midrule
\multirow{2}{*}{1} 
  & Q-EANet & \textbf{0.19}/{0.17} & \textbf{0.22}/{0.19} & 0.28/0.26                & 0.22/0.21                & \cellcolor{gray!20}\textbf{0.11}/\textbf{0.09} & 0.15/0.13                \\
  & SAIL    & 0.20/\textbf{0.16}         & 0.23/\textbf{0.18}         & \cellcolor{gray!20}\textbf{0.26}/\textbf{0.24} & \cellcolor{gray!20}\textbf{0.21}/\textbf{0.19} & 0.12/0.10                & \cellcolor{gray!20}\textbf{0.14}/\textbf{0.12} \\
\midrule
\multirow{2}{*}{2} 
  & Q-EANet & \textbf{0.30}/{0.29} & 0.36/0.34                  & 0.42/0.44                & 0.34/0.37                & \cellcolor{gray!20}\textbf{0.18}/\textbf{0.18} & 0.23/0.23                \\
  & SAIL    & 0.31/\textbf{0.27}         & \cellcolor{gray!20}\textbf{0.34}/\textbf{0.31} & \cellcolor{gray!20}\textbf{0.39}/\textbf{0.40} & \cellcolor{gray!20}\textbf{0.32}/\textbf{0.34} & 0.19/0.19                & \cellcolor{gray!20}\textbf{0.21}/\textbf{0.21} \\
\midrule
\multirow{2}{*}{3} 
  & Q-EANet & 0.42/0.45                  & 0.49/0.50                  & 0.56/0.62                & 0.47/0.56                & 0.29/0.32                & 0.35/0.38                \\
  & SAIL    & \cellcolor{gray!20}\textbf{0.39}/\textbf{0.40} & \cellcolor{gray!20}\textbf{0.46}/\textbf{0.46} & \cellcolor{gray!20}\textbf{0.52}/\textbf{0.56} & \cellcolor{gray!20}\textbf{0.44}/\textbf{0.51} & \cellcolor{gray!20}\textbf{0.27}/\textbf{0.29} & \cellcolor{gray!20}\textbf{0.32}/\textbf{0.34} \\
\midrule
\multirow{2}{*}{4} 
  & Q-EANet & 0.55/0.63                  & 0.61/0.69                  & 0.74/0.82                & 0.63/0.78                & 0.39/0.46                & 0.46/0.54                \\
  & SAIL    & \cellcolor{gray!20}\textbf{0.51}/\textbf{0.57} & \cellcolor{gray!20}\textbf{0.57}/\textbf{0.63} & \cellcolor{gray!20}\textbf{0.68}/\textbf{0.74} & \cellcolor{gray!20}\textbf{0.58}/\textbf{0.71} & \cellcolor{gray!20}\textbf{0.36}/\textbf{0.42} & \cellcolor{gray!20}\textbf{0.42}/\textbf{0.49} \\
\midrule
\multirow{2}{*}{5} 
  & Q-EANet & 0.70/0.83                  & 0.78/0.89                  & 0.95/1.08                & 0.81/1.04                & 0.53/0.64                & 0.59/0.72                \\
  & SAIL    & \cellcolor{gray!20}\textbf{0.64}/\textbf{0.75} & \cellcolor{gray!20}\textbf{0.72}/\textbf{0.81} & \cellcolor{gray!20}\textbf{0.87}/\textbf{0.97} & \cellcolor{gray!20}\textbf{0.74}/\textbf{0.94} & \cellcolor{gray!20}\textbf{0.48}/\textbf{0.58} & \cellcolor{gray!20}\textbf{0.54}/\textbf{0.65} \\
\midrule
\multirow{2}{*}{6} 
  & Q-EANet & 0.86/1.14                  & 0.96/1.15                  & 1.13/1.64                & 0.97/1.37                & 0.61/0.89                & 0.70/0.99                \\
  & SAIL    & \cellcolor{gray!20}\textbf{0.80}/\textbf{1.01} & \cellcolor{gray!20}\textbf{0.90}/\textbf{1.08} & \cellcolor{gray!20}\textbf{1.10}/\textbf{1.41} & \cellcolor{gray!20}\textbf{0.94}/\textbf{1.28} & \cellcolor{gray!20}\textbf{0.61}/\textbf{0.78} & \cellcolor{gray!20}\textbf{0.69}/\textbf{0.88} \\
\bottomrule
\end{tabular}}
\label{table_4}
\end{table*}

\begin{table*}[htbp]
  \centering
  \caption{Model-agnostic worst-case performance comparison on the nuScenes dataset, reported in $\text{minADE}_5$ / $\text{minFDE}_5$. The Top 1\% to 5\% subsets are dynamically defined for each model based on its own worst-performing samples, ranking by $\text{minFDE}_5$. The best results are in bold, and the second best are underlined.}
    \begin{tabular}{l|c|ccccc|c}
    \toprule
    Model & Venue & Top 1\% & Top 2\% & Top 3\% & Top 4\% & Top 5\% & All \\
    \midrule
    PGP \cite{deo2022multimodal} & CoRL 2021 & 8.86/21.92 & 7.21/17.90 & 6.24/15.68 & 5.52/13.77 & 5.02/12.44 & 1.28/2.52 \\
    Q-EANet \cite{chen2024q-qeanet} & ITS 2024 & \underline{7.55}/\underline{18.78} & \underline{6.15}/\underline{15.58} & \underline{5.44}/\underline{13.76} & \underline{4.94}/\underline{12.49} & \underline{4.55}/\underline{11.49} & 1.20/2.45 \\
    LAformer \cite{liu2024laformer} & CVPR 2024 & 8.19/19.03 & 6.73/15.81 & 5.89/13.90 & 5.33/12.60 & 4.90/11.61 & 1.19/\underline{2.42} \\
    UniTraj (MTR) \cite{feng2024unitraj} & ECCV 2024 & 7.84/21.69 & 6.44/18.06 & 5.69/15.95 & 5.18/14.49 & 4.78/13.37 & \textbf{1.15}/2.61 \\
    \midrule
    \rowcolor{gray!20} \textbf{SAIL} & - & \textbf{6.84}/\textbf{17.21} & \textbf{5.80}/\textbf{14.57} & \textbf{5.05}/\textbf{12.88} & \textbf{4.58}/\textbf{11.68} & \textbf{4.22}/\textbf{10.72} & \underline{1.18}/\textbf{2.42} \\
    \bottomrule
    \end{tabular}%
  \label{tab5}%
\end{table*}%

{\subsubsection{Quantitative Results on Worst-Case Samples}}
To reduce the bias that may arise from defining hard cases using a single fixed baseline, we further adopt a model-agnostic worst-case evaluation protocol. Specifically, for each model, we rank all test samples according to its own $\text{minFDE}_5$ values and construct the Top 1\% to Top 5\% hardest subsets accordingly. This protocol focuses on the upper tail of each model's error distribution and therefore provides a more reliable assessment of robustness under highly challenging scenarios. As shown in Table~\ref{tab5}, SAIL consistently achieves the best performance across all worst-case subsets in both $\text{minADE}_5$ and $\text{minFDE}_5$. On the Top 1\% hardest samples, SAIL attains 6.84/17.21, outperforming the strongest competing method, by 9.4\% in $\text{minADE}_5$ and 8.4\% in $\text{minFDE}_5$. More importantly, this advantage remains stable as the subset size expands from Top 2\% to Top 5\%, where SAIL continues to rank first with clear margins over all baselines. These results indicate that the proposed method not only improves average prediction quality, but also more effectively suppresses extreme failures in the most difficult and safety-critical cases.

{\subsubsection{Quantitative Results on the Full Dataset}}
This section presents the overall quantitative comparison of SAIL with leading baselines on the nuScenes and ETH/UCY benchmarks. On the large-scale and complex nuScenes dataset, the results on the full dataset, as shown in Table \ref{table_6}, demonstrate strong performance across multiple key metrics. SAIL achieves the best {minFDE}$_1$ of 6.45, representing a 7.7\% improvement over the second-best approach. Furthermore, it attains a leading minADE$_5$ of 1.18 and the lowest MR$_5$ of 0.50. This comprehensive success indicates that our model not only predicts the overall trajectory shape more accurately but is also more reliable in forecasting the final crucial endpoint and avoiding significant prediction failures, which are vital capabilities for real-world driving scenarios. This strong performance is further corroborated on the ETH/UCY datasets, which feature diverse pedestrian dynamics. As shown in Table \ref{table_7}, SAIL achieves the best average error across all five scenes. This corresponds to a significant reduction of 5.5\% in average minADE compared to the strongest competing method. This consistent superiority demonstrates that our model performs well not only in vehicle-centric scenarios like nuScenes but is also highly effective in pedestrian-focused environments. This validates the robust generalization and strong predictive capability of our SAIL model.

\begin{table}[htbp]
\centering
\caption{Quantitative evaluation of trajectory prediction performance on the full nuScenes dataset, with a prediction horizon of 6 s. Light gray background indicates the performance of our model.}
\begin{tabular}{l|c|cccccc}
\toprule
Model   & Venue      & minADE$_{5}$ & minADE$_{10}$ & minFDE$_{1}$ & minFDE$_{5}$ & minFDE$_{10}$ & MR$_{5}$ \\
\midrule
Trajectron++ \cite{salzmann2020trajectron++}    & ECCV 2020 & 1.88 & 1.51 & 9.52 & 5.63 & - & 0.70 \\
LaPred \cite{kim2021lapred}                     & CVPR 2021 & 1.47 & 1.12 & 8.12 & 3.37 & 2.39 & \uline{0.53} \\
MHA-JAM \cite{messaoud2021trajectory}                       & IV 2021 & -    & 1.18 & 9.62 & 3.72 & \uline{2.21} & 0.64 \\
GoHome \cite{gilles2022gohome}                  & ICRA 2022 & 1.42 & 1.15 & \uline{6.99} & - & - & 0.57 \\
ContextVAE \cite{xu2023context-VAE}             & RAL 2023  & 1.59 & -    & 8.24 & 3.28 & - & - \\
AFormer-FLN \cite{xu2024adapting-AFormer}       & CVPR 2024 & 1.83 & 1.32 & -    & 3.78 & 2.86 & - \\
EMSIN \cite{ren2024emsin}                       & TFS 2024  & 1.77 & 1.28 & 9.06 & 3.56 & - & 0.54 \\
WAKE \cite{wang2025wake}                        & TPAMI 2025& \uline{1.24} & 1.09 & 7.02 & \uline{2.96} & 2.37 & 0.55 \\
SeFlow \cite{zhang2025seflow}                   & ECCV 2025 & 1.38 & \textbf{0.98} & 7.89 & - & - & 0.60 \\
\midrule
\rowcolor{gray!20} 
SAIL & - & \textbf{1.18} & \uline{1.03} & \textbf{6.45} & \textbf{2.42} & \textbf{1.99} & \textbf{0.50} \\
\bottomrule
\end{tabular}
\label{table_6}
\end{table}

\begin{table*}[htbp]
\centering
\caption{Quantitative evaluation of trajectory prediction performance on the ETH/UCY dataset across all samples. Light gray background indicates the performance of our model.}
\begin{tabular}{l|c|ccccc|c} 
\toprule
Model          & Venue           & ETH                     & HOTEL                   & UNIV                    & ZARA1                   & ZARA2                   & AVG                      \\ 
\midrule
PECNet \cite{mangalam2020not}         & ECCV 2020          & 0.54/0.87               & 0.18/0.24               & 0.22/0.39               & 0.17/0.30               & 0.35/0.60               & 0.29/0.48                \\ 
AgentFormer \cite{yuan2021agentformer}     & ICCV 2021          & 0.45/0.75               & 0.14/0.22               & 0.25/0.45               & 0.18/0.30               & 0.14/0.24               & 0.23/0.39                \\ 
Trajectron++ \cite{salzmann2020trajectron++}   & ECCV 2020         & 0.39/0.83               & 0.12/0.21               & {0.20}/0.44               & \uline{0.15}/0.33               & \textbf{0.11}/0.25               & {0.19}/0.41                \\ 

NPSN \cite{bae2022non}       & CVPR 2022             & 0.36/0.59 & 0.16/0.25               & 0.23/{0.39}               & 0.18/0.32               & 0.14/0.25               & 0.21/0.36                \\ 
MID \cite{gu2022stochastic}       & CVPR 2022              & 0.39/0.66               & 0.13/0.22 & 0.22/0.45               & 0.17/0.30               & 0.13/0.27        & 0.21/0.38                \\ 
TUTR \cite{shi2023trajectory}      & ICCV 2023              & 0.40/0.61               & 0.11/0.18 & 0.23/0.42               & 0.18/0.34               & 0.13/0.25        & 0.21/0.36                \\ 
PPT \cite{lin2024progressive}       & ECCV 2024              & \uline{0.36}/{0.51} & \uline{0.11}/\uline{0.15} & 0.22/0.40               & 0.17/{0.30}               & 0.12/{0.21}        & 0.20/{0.31} \\ 
UniEdge \cite{li2025unified}       & TCSVT 2025              & {0.36}/\uline{0.46} & {0.11}/{0.17}      & \uline{0.19}/\textbf{0.28}       & \textbf{0.14}/{\uline{0.27}}         & {0.12}/\textbf{0.16}      & \uline{0.18}/\uline{0.27} \\
\midrule
 \rowcolor{gray!20} SAIL       & -          & \textbf{0.30}/\textbf{0.43} & \textbf{0.09}/\textbf{0.12} & \textbf{0.18}/\uline{0.35} & {0.16}/\textbf{0.25} & \uline{0.12}/\uline{0.19} & \textbf{0.17}/\textbf{0.27} \\ 
\bottomrule
\end{tabular}
\label{table_7}%
\end{table*}

\begin{table*}[htbp]
\centering
\caption{Ablation analysis of key components within our method on the nuScenes benchmark. Each row from A to E represents a variant of our full model with one key component removed. AGTA: Attribute-Guided Trajectory Augmentation, ADP: Attribute Disentanglement and Prediction, AMCL: Adaptive Momentum Contrastive Learning, EFC: Evolving Feature Clustering, FDCL: Focused Decoupled Contrastive Learning. \textbf{Bold} text represents the best results.}
\resizebox{\linewidth}{!}{
\begin{tabular}{c|ccccc|ccccccc}
\toprule
\multirow{2}[2]{*}{Model} & \multicolumn{5}{c|}{Components} & \multicolumn{7}{c}{Performance (minADE/minFDE)} \\
\cmidrule(lr){2-6} \cmidrule(lr){7-13}
 & AGTA & ADP & AMCL & EFC & FDCL & Top 1\% & Top 2\% & Top 3\% & Top 4\% & Top 5\% & Rest & All \\
\midrule
A & $\times$ & $\checkmark$ & $\checkmark$ & $\checkmark$ & $\checkmark$  & 1.30/1.88 & 1.05/1.55 & 0.88/1.35 & 0.79/1.24 & 0.73/1.16 & 0.19/0.23 & 0.22/0.27\\
B & $\checkmark$ & $\times$ & $\checkmark$ & $\checkmark$ & $\checkmark$ & 1.48/2.10 & 1.15/1.70 & 0.95/1.48 & 0.85/1.35 & 0.79/1.26 & 0.21/0.24 & 0.24/0.28 \\
C & $\checkmark$ & $\checkmark$ & $\times$ & $\checkmark$ & $\checkmark$ & 1.25/1.83 & 1.01/1.51 & 0.84/1.31 & 0.76/1.21 & 0.70/1.13 & 0.18/0.22 & 0.21/0.26 \\
D & $\checkmark$ & $\checkmark$ & $\checkmark$ & $\times$ & $\checkmark$ & 1.15/1.72 & 0.92/1.41 & 0.75/1.22 & 0.68/1.13 & 0.63/1.06 & 0.18/0.21 & 0.20/0.25 \\
E & $\checkmark$ & $\checkmark$ & $\checkmark$ & $\checkmark$ & $\times$ & 1.22/1.80 & 0.98/1.49 & 0.81/1.29 & 0.73/1.19 & 0.68/1.11 & 0.19/0.22 & 0.21/0.26 \\
\midrule
\rowcolor{gray!20}
F & $\checkmark$ & $\checkmark$ & $\checkmark$ & $\checkmark$ & $\checkmark$ & \textbf{1.02}/\textbf{1.58} & \textbf{0.81}/\textbf{1.30} & \textbf{0.65}/\textbf{1.12} & \textbf{0.60}/\textbf{1.05} & \textbf{0.56}/\textbf{0.98} & \textbf{0.16}/\textbf{0.19} & \textbf{0.18}/\textbf{0.23} \\
\bottomrule
\end{tabular}}
\label{table_8}
\end{table*}

\subsection{Ablation Analysis}

Table \ref{table_8} presents a systematic ablation study to evaluate the contribution of each component in our framework. Due to dependencies between certain modules, the ablation settings are designed with minimal fallback implementations to preserve the functionality of the remaining components while removing the target module under study. Specifically, for {Model A (w/o AGTA)}, the attribute-guided augmentation is replaced by random augmentation from the same augmentation pool, where one augmentation strategy is randomly selected for each sample without using attribute guidance. For {Model B (w/o ADP)}, the attribute-supervised auxiliary loss is removed. For {Model C (w/o AMCL)}, the adaptive momentum contrastive loss is removed. For {Model D (w/o EFC)}, the evolving clustering process is replaced by static clustering, where pseudo-labels are generated once after the warm-up stage and then kept fixed during the remaining training. This fallback is used to preserve the pseudo-label source required by FDCL after removing the dynamic update mechanism. For {Model E (w/o FDCL)}, the focused decoupled contrastive loss is removed.

Under these settings, the results confirm that the complete framework (Model F) attains the best results across all evaluation metrics, demonstrating a powerful synergy between its modules. Each subsequent model A-E represents a variant of the full model with one key component removed. The results show that the attribute-guided modules, {Model A (w/o AGTA)} and {Model B (w/o ADP)}, are particularly important to the success of the framework. Removing either of these modules leads to the most significant performance degradation. For instance, Model B (w/o ADP) sees its minFDE on the Top 1\% of samples increase by a substantial 32.9\% compared to the full model. This suggests that explicitly modeling and supervising trajectory attributes is crucial for capturing complex long-tail behaviors. The contrastive learning modules also prove to be vital. Model C (w/o AMCL), which lacks the core unsupervised representation learning stage, exhibits a sharp decline in performance, highlighting the importance of learning a discriminative feature space. Furthermore, Model E (w/o FDCL) and Model D (w/o EFC) show noticeable performance drops, confirming that FDCL's ability to focus on hard-positive samples and EFC's dynamic pseudo-labeling are crucial for refining class boundaries and providing high-quality supervision. Overall, the ablation results show that each component contributes positively to the final framework, while the best performance is achieved when all modules are used together. This demonstrates that the advantage of SAIL is the result of a coordinated design, in which attribute-guided learning improves long-tail awareness, contrastive objectives enhance representation quality, and dynamic pseudo-label refinement further strengthens class discrimination in challenging scenarios.

\begin{table}[htbp]
\centering
\caption{{Jaccard similarity between attribute-specific long-tail subsets on the nuScenes validation set.} The overlap between subsets identified by Prediction Error, Collision Risk, and State Complexity is measured under different tail thresholds, from the Top 20\% to the Top 5\% subsets.}
\resizebox{0.8\textwidth}{!}{
\begin{tabular}{@{}ccccc@{}}
\toprule
\multirow{2}[2]{*}{{Attribute Pair}} & \multicolumn{4}{c}{{Jaccard Similarity }} \\ \cmidrule(l){2-5} 
 & {Top 20\%} & {Top 15\%} & {Top 10\%} & {Top 5\%} \\ \midrule
Prediction Error vs. Collision Risk  & 10.7\% & 8.3\% & 5.2\% & 1.4\% \\ 
Prediction Error vs. State Complexity & 13.0\% & 10.3\% & 6.8\% & 4.8\% \\
Collision Risk vs. State Complexity & 17.0\% & 14.1\% & 11.4\% & 5.7\% \\
\bottomrule
\end{tabular}%
}
\label{tab9}
\end{table}

\begin{figure}[pos=t]
  \centering
  \includegraphics[width=0.8\linewidth]{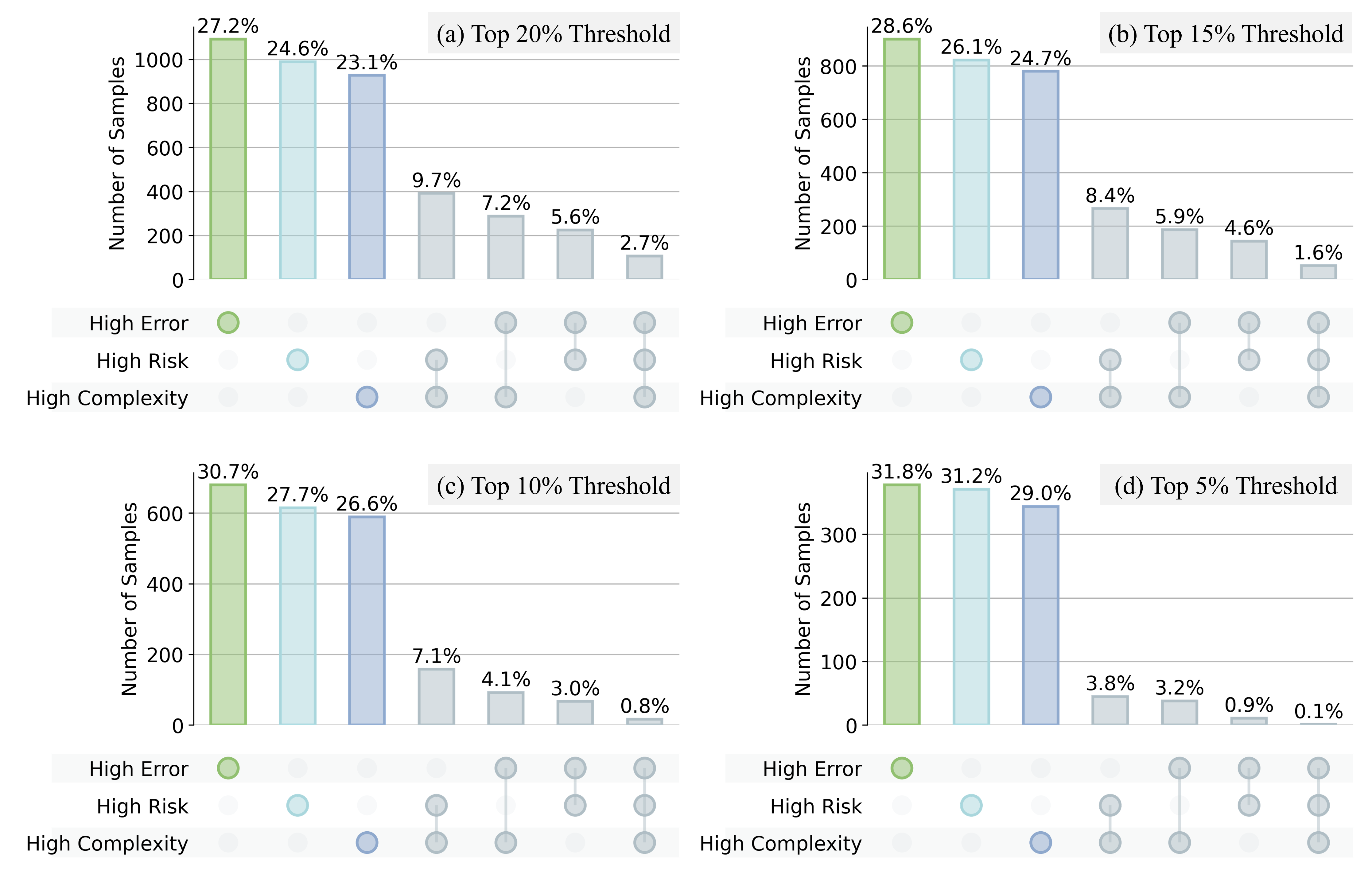}
   \caption{{UpSet visualization of intersections among attribute-specific long-tail subsets on the nuScenes validation set.} Subfigures (a) to (d) correspond to the Top 20\%, Top 15\%, Top 10\%, and Top 5\% tail thresholds, respectively. In each subfigure, the bars indicate the number of samples in each subset intersection, while the connected dots denote the corresponding combination of Prediction Error, Collision Risk, and State Complexity. The percentages above the bars represent the proportion of each intersection among all samples selected under the given threshold.}
   \label{fig5}
\end{figure}

\subsection{Validation of the Long-Tail Attributes}
To further examine the properties of the evaluated long-tail scenarios and validate the proposed three-dimensional taxonomy consisting of Prediction Error, Collision Risk, and State Complexity, we analyze the overlap among their corresponding sample subsets on the validation set. Table~\ref{tab9} reports the Jaccard similarity between attribute-specific long-tail subsets under progressively stricter tail thresholds. A clear decreasing trend can be observed: as the analysis moves from the Top 20\% to the Top 5\% subsets, the overlap between different attributes steadily decreases. This result indicates that the samples identified by the three attributes become increasingly distinct as the focus shifts toward more extreme tail cases.
The complementary nature of these attributes is further illustrated by the visualization in Figure~\ref{fig5}, where most of the most challenging samples are associated with a single attribute rather than their intersections. Notably, the overlap among all three attribute pairs shows a consistent downward trend as the tail threshold becomes stricter, indicating that different attribute-specific scenarios become increasingly differentiated in the more extreme tail regions. Overall, these results show that Prediction Error, Collision Risk, and State Complexity are not strictly independent, but they are not redundant either. Instead, they capture complementary aspects of long-tail trajectory prediction, and together provide a more comprehensive characterization of challenging scenarios in the dataset.

\subsection{Inference Efficiency Analysis}
\begin{figure}[pos=t]
  \centering
  \includegraphics[width=0.7\linewidth]{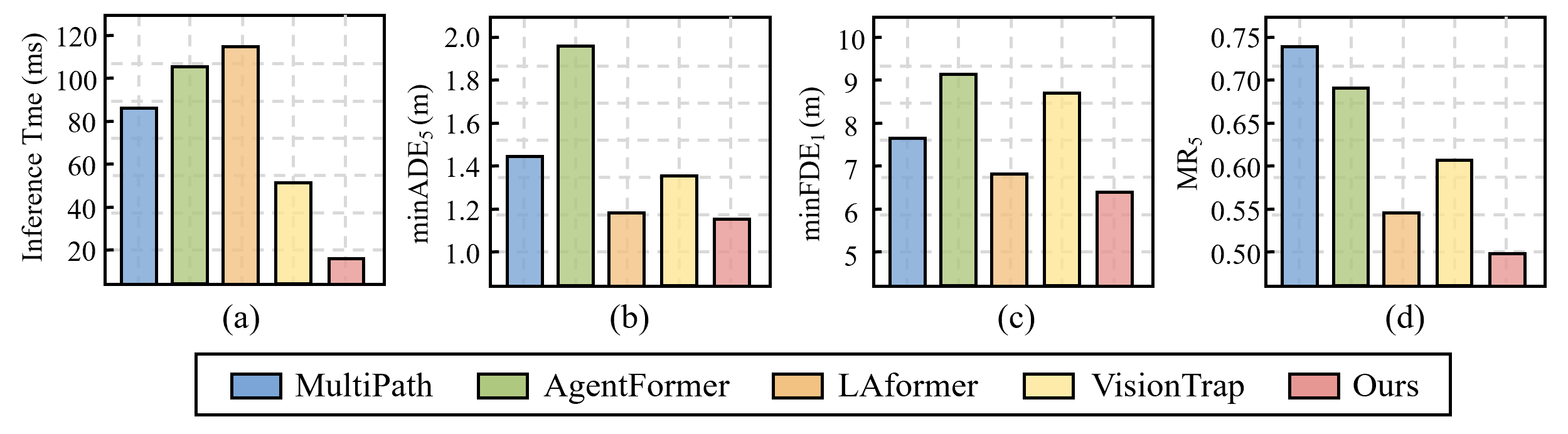}
   \caption{Visualization of comparison results for inference time and error metrics on the nuScenes dataset. Panel (a) shows inference time, panel (b) depicts \(\text{minADE}_{5}\), panel (c) illustrates \(\text{minFDE}_{1}\), and panel (d) presents \(\text{MR}_{5}\).}
   \label{fig6}
\end{figure}

To evaluate the trade-off between computational efficiency and predictive accuracy, we compare the end-to-end inference time and performance of SAIL against several state-of-the-art baselines on the nuScenes dataset. For consistency with prior work, we report the average inference time for predicting 12 agents across the nuScenes test set. To ensure a fair hardware comparison, all latency measurements are conducted on a single NVIDIA RTX 3090 GPU. As shown in Table \ref{tab10}, SAIL achieves an end-to-end inference latency of only 18 ms, which is 66.0\% lower than that of the fastest competing baseline. This high efficiency mainly comes from our asymmetric design: computationally intensive components, including the attribute prediction supervision branch, trajectory augmentation, and contrastive learning modules, are used only during training and introduce no additional overhead at test time. During inference, these auxiliary branches are removed, leaving a streamlined pipeline for attribute disentanglement and trajectory prediction.
Importantly, this efficiency is achieved together with strong predictive performance. SAIL attains the best results across all evaluated metrics, improving upon the second-best method by 6.4\% in the safety-critical $\text{minFDE}_1$ metric while also achieving the best $\text{minADE}_5$ and $\text{MR}_5$. Figure \ref{fig6} further illustrates the balance between inference time and prediction accuracy. These results show that SAIL achieves both low runtime latency and strong long-tail prediction capability, making it well-suited for real-world autonomous driving systems with strict efficiency requirements.

\begin{table}[htbp]
    \centering
    \caption{Comparative analysis of computational efficiency and predictive accuracy on the nuScenes dataset. We compare SAIL against SOTA baselines to demonstrate the trade-off between speed and performance. \textbf{Bold} and \uline{underlined} text represent the best and second-best results, respectively.}
\begin{tabular}{l|cccc}
        \toprule
        Model & Inference Time (ms) & $\text{minADE}_{5}$ & $\text{minFDE}_{1}$ & $\text{MR}_{5}$ \\
        \midrule
        MultiPath \cite{chai2019multipath} & 87 & 1.44 & 7.69 & 0.74 \\
        AgentFormer \cite{yuan2021agentformer} & 107 & 1.97 & 9.12 & 0.69 \\
        LAformer \cite{liu2024laformer} & 115 & \uline{1.19} & \uline{6.89} & \uline{0.54} \\
        VisionTrap \cite{moon2024visiontrap} & \uline{53} & 1.36 & 8.72 & 0.61 \\
        \midrule
 \rowcolor{gray!20}
        \textbf{SAIL} & \textbf{18} & \textbf{1.18} & \textbf{6.45} & \textbf{0.50} \\
        \bottomrule
    \end{tabular}
\label{tab10}%
\end{table}

\subsection{Analysis of the Clustering Strategy}
\subsubsection{Sensitivity to the Number of Clusters}
The number of clusters $C$ is a critical hyperparameter that determines the granularity of the pseudo-labels. A small $C$ may merge distinct long-tail patterns, while a large $C$ could fragment similar patterns into separate clusters, leading to noisy supervision. To find the optimal balance, we perform a sensitivity analysis on the nuScenes dataset, with results shown in Table \ref{tab11}. The performance improves as $C$ increases from 2 to 5, peaking at $C=5$. Beyond this point, increasing $C$ leads to a slight degradation in performance, likely due to overfitting on finer-grained but less generalizable sub-patterns. Therefore, we set $C=5$ as the optimal number of clusters in all our experiments, as it provides the best balance between capturing diverse trajectory behaviors and maintaining robust generalization.

\begin{table}[htbp]
\centering
\caption{Performance (minADE/minFDE) on nuScenes for varying numbers of clusters ($C$) in our EFC module.}
\begin{tabular}{c|ccccccc}
\toprule
$C$ & Top 1\% & Top 2\% & Top 3\% & Top 4\% & Top 5\% & Rest & All \\
\midrule
2 & 1.35/1.95 & 1.08/1.60 & 0.90/1.40 & 0.82/1.30 & 0.76/1.22 & 0.20/0.24 & 0.23/0.27 \\
3 & 1.21/1.80 & 0.98/1.49 & 0.82/1.29 & 0.74/1.20 & 0.68/1.12 & 0.18/0.22 & 0.21/0.25 \\
4 & 1.10/1.70 & 0.88/1.38 & 0.76/1.20 & 0.67/1.12 & 0.62/1.05 & 0.17/0.20 & 0.19/0.24 \\
\rowcolor{gray!20}
\textbf{5} & \textbf{1.02}/\textbf{1.58} & \textbf{0.81}/\textbf{1.30} & \textbf{0.65}/\textbf{1.12} & \textbf{0.60}/\textbf{1.05} & \textbf{0.56}/\textbf{0.98} & \textbf{0.16}/\textbf{0.19} & \textbf{0.18}/\textbf{0.23} \\
6 & 1.05/1.62 & 0.84/1.34 & 0.69/1.16 & 0.62/1.08 & 0.58/1.01 & 0.16/0.20 & 0.18/0.23 \\
7 & 1.12/1.73 & 0.90/1.42 & 0.75/1.24 & 0.68/1.15 & 0.63/1.08 & 0.17/0.21 & 0.20/0.25 \\
\bottomrule
\end{tabular}
\label{tab11}
\end{table}

\begin{figure*}[pos=t]
  \centering
\includegraphics[width=0.8\linewidth]{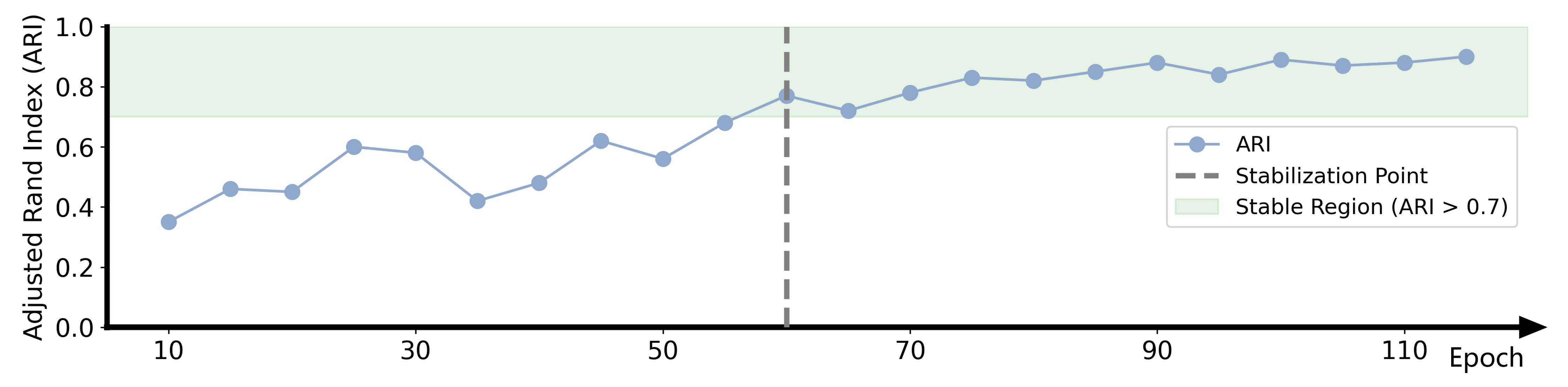} 
  \caption{Stability analysis of the Evolving Feature Clustering (EFC) process. The Adjusted Rand Index (ARI) between consecutive clustering updates is plotted against training epochs on the nuScenes dataset.}
  \label{fig7} 
\end{figure*}

\subsubsection{Stability of the Clustering Process}
To validate the stability of our EFC strategy throughout the training process, we analyze the consistency of the pseudo-label assignments over time. We measure the stability of our clustering process using the Adjusted Rand Index (ARI). The ARI quantifies the similarity between two data clusterings; a higher ARI indicates greater consistency. We compute the ARI between the pseudo-label assignments of consecutive clustering updates. As shown in Figure \ref{fig7}, the ARI is initially low during the early stages of training, as the feature space is still rapidly evolving. However, after a warm-up period, the ARI value quickly rises and stabilizes at a high level, indicating that the cluster assignments become highly consistent. This result confirms that our EFC strategy produces a stable and reliable supervisory signal, successfully avoiding the potential pitfalls of noisy, fluctuating pseudo-labels.

\begin{figure*}[pos=t]
  \centering
\includegraphics[width=0.7\textwidth]{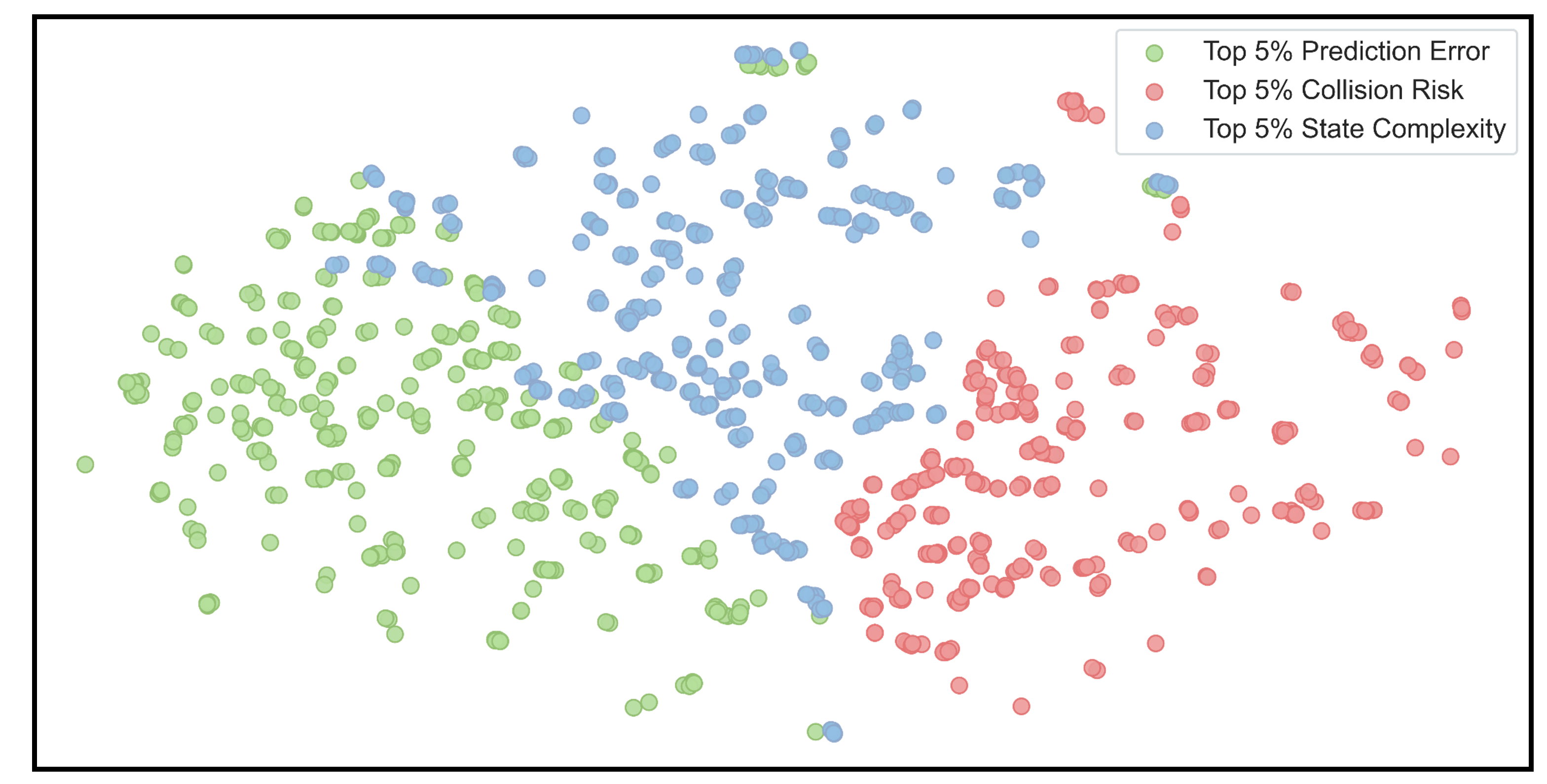} 
  \caption{t-SNE visualization of samples drawn from the Top 5\% subsets under the three-dimensional long-tail attributes of prediction error, collision risk, and state complexity. The three groups exhibit relatively distinct yet partially overlapping manifold structures, indicating that they capture different but correlated aspects of long-tail driving scenarios.}
  \label{fig8} 
\end{figure*}

\begin{figure*}[pos=t]
  \centering
\includegraphics[width=0.85\textwidth]{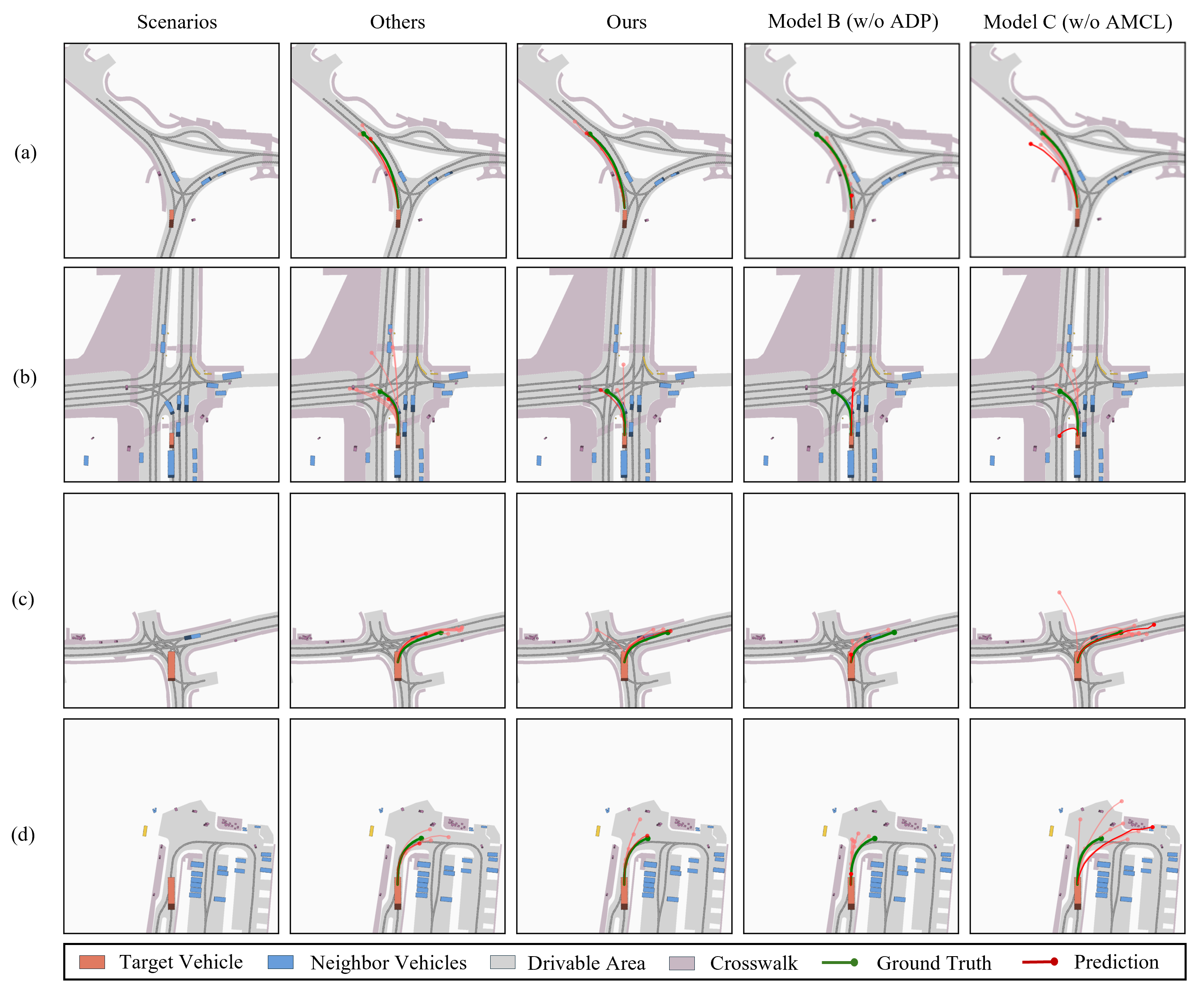} 
      \caption{Qualitative results of long-tail trajectory prediction across diverse high-curvature driving scenarios on the nuScenes dataset. Panels (a) and (b) depict left-turn maneuvers, while panels (c) and (d) illustrate right-turn maneuvers. The others represent baseline model \cite{chen2024q-qeanet} predictions. Model B and Model C represent the variants of our model. The red line depicts the highest-probability trajectory, whereas the light red lines illustrate the multimodal predictions.}
  \label{fig9} 
\end{figure*}

\begin{figure*}[pos=t]
  \centering
\includegraphics[width=0.85\textwidth]{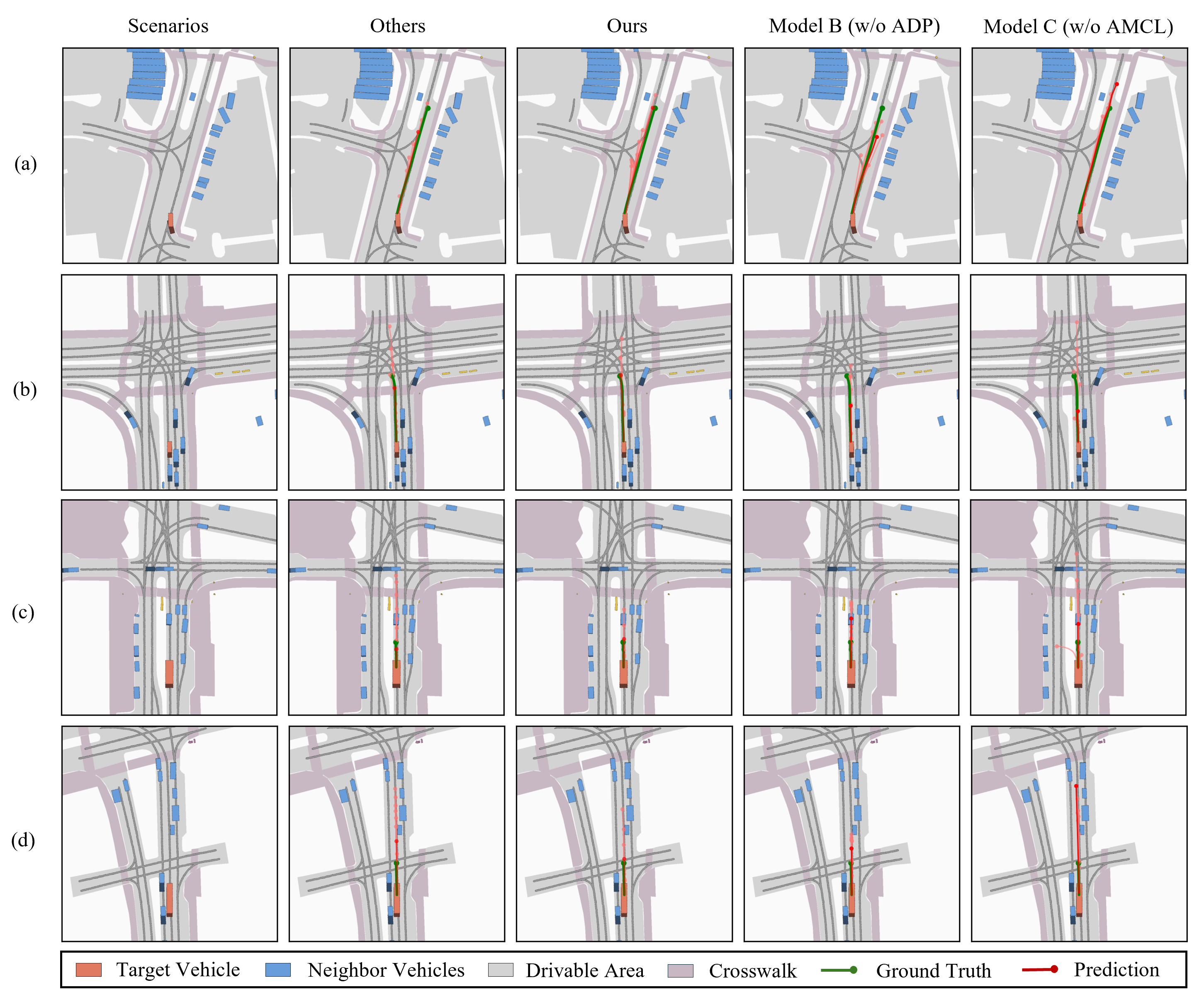} 
  \caption{Qualitative results of long-tail trajectory prediction across various acceleration and deceleration scenarios on the nuScenes dataset. Panels (a) and (b) depict acceleration maneuvers, while panels (c) and (d) illustrate deceleration maneuvers. The others represent baseline model \cite{chen2024q-qeanet} predictions. Model B and Model C represent the variants of our model. The red line depicts the highest-probability trajectory, whereas the light red lines illustrate the multimodal predictions. }
  \label{fig10} 
\end{figure*}

\begin{figure*}[pos=h]
  \centering
\includegraphics[width=0.90\textwidth]{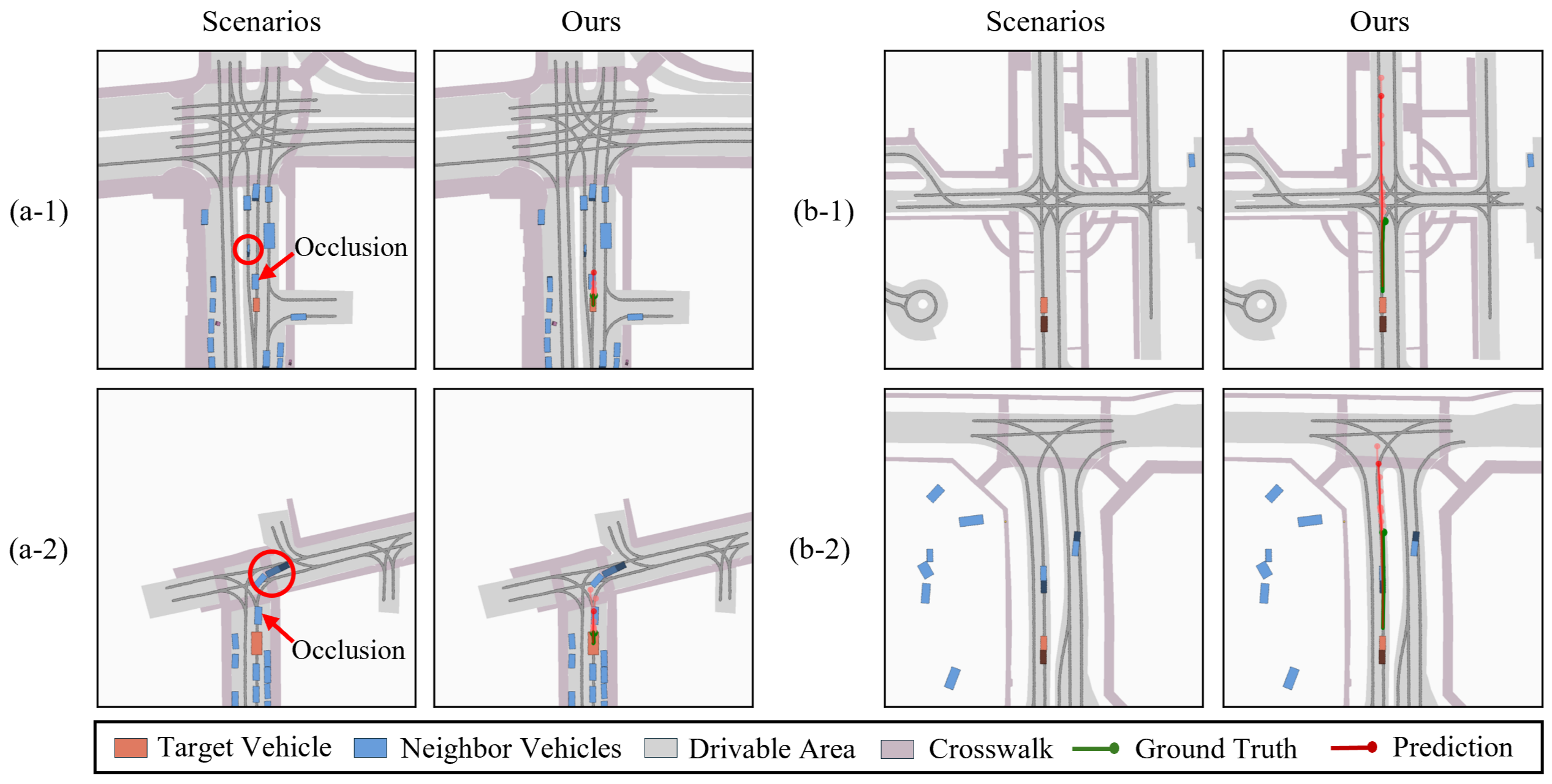} 
  \caption{Representative failure cases under extreme long-tail scenarios. (a) Failure cases caused by severe visual occlusion. (b) Failure cases at open intersections without explicit traffic signal information. The red line depicts the highest-probability trajectory, whereas the light red lines illustrate the multimodal predictions. }
  \label{fig11} 
\end{figure*}

\subsection{Qualitative Results}

\subsubsection{Visualization of Disentangled Feature Space}
To further examine the structural relationship among the three-dimensional attributes of long-tail samples, we project the learned representations into a two-dimensional space using t-SNE, as shown in Figure~\ref{fig8}. Each point corresponds to a sample drawn from the Top 5\% subset under prediction error, collision risk, or state complexity. A notable observation is that the three groups do not collapse into a single mixed distribution, but instead exhibit relatively distinct manifold organizations. This indicates that the proposed three-dimensional attribute space is not merely assigning different names to the same set of difficult samples. Rather, each attribute emphasizes a different structural aspect of long-tail scenarios, suggesting that prediction error, collision risk, and state complexity correspond to meaningfully different forms of rarity and challenge in real-world driving data.

Furthermore, the separation is not absolute, and partial overlap remains among the three groups. This observation is important, because it implies that the three attributes are not independent in a strict sense, but are linked through shared underlying difficulty. In practice, genuinely critical driving samples often simultaneously exhibit elevated uncertainty, increased safety risk, and greater behavioral complexity, even though one attribute may be more dominant than the others in a given case. Therefore, the t-SNE visualization supports a more nuanced conclusion: the proposed three-dimensional attributes are neither redundant nor fully separable, but complementary dimensions that jointly characterize the heterogeneity of long-tail scenarios. This qualitative evidence further justifies the use of a multi-attribute framework, rather than any single-dimensional definition, for identifying rare and critical driving samples.

\subsubsection{Trajectory Visualization in Long-Tail Scenarios}
To further validate the accuracy of our model in predicting long-tail trajectories, we visualize multimodal prediction results on nuScenes across challenging scenarios, comparing our SAIL model with a baseline model \cite{chen2024q-qeanet} and two of its key ablation variants, Model B (w/o ADP) and Model C (w/o AMCL).

\textbf{(1) High-curvature Turning Scenarios.} Figure~\ref{fig9} illustrates high-curvature turns, a classic long-tail scenario where an agent's future path deviates significantly from its historical heading. The compared models exhibit distinct failure modes. The baseline model struggles to fully grasp the complexity of this maneuver, often resulting in inaccurate endpoint predictions even when the general turning direction is captured. This indicates a rudimentary understanding of the required trajectory geometry. The ablation variants reveal more specific weaknesses. Model B (w/o ADP), deprived of explicit attribute information, demonstrates a more fundamental failure: it seems unable to recognize the latent intent hidden within the subtle cues of this rare maneuver, instead defaulting to a simplistic forward projection. Model C (w/o AMCL), lacking the robust feature representations from our adaptive contrastive learning, suffers from high uncertainty and randomness, generating a scattered and unreliable set of trajectories. In contrast, our full SAIL model excels. Aided by the ADP's contextual understanding and the discriminative features learned by AMCL, SAIL not only predicts the ground-truth path with pinpoint accuracy but also generates other plausible, high-quality alternatives.

\textbf{(2) Abrupt Speed Change Scenarios.} Figure~\ref{fig10} shows prediction results in scenarios characterized by abrupt, non-linear speed changes, another typical manifestation of long-tail trajectory behavior. In these cases, the baseline and ablation models commonly exhibit prediction lag, particularly under hard deceleration, where their predicted trajectories overshoot the true stopping position. This indicates that they tend to assume a more common smooth braking profile and thus fail to respond promptly to rare but critical speed transitions. Model B (w/o ADP) is less capable of distinguishing these extreme dynamic patterns from ordinary velocity fluctuations, while Model C (w/o AMCL) shows weaker stability in capturing the timing and magnitude of sudden motion changes. By contrast, SAIL more accurately tracks both acceleration and deceleration trends, especially in cases involving abrupt braking or rapid velocity shifts. This qualitative advantage suggests that modeling multi-dimensional long-tail attributes helps the framework better identify rare dynamic intent, while the contrastive representation learning further improves separability for these hard cases. As a result, SAIL demonstrates stronger robustness in scenarios where non-linear motion change is the key source of prediction difficulty.

\subsection{Discussion of Failure Cases}
Despite the strong overall performance of SAIL across diverse long-tail scenarios, prediction errors may still arise in a few extreme cases where critical contextual cues are unavailable or highly ambiguous. Figure \ref{fig11} presents two representative types of such failure cases. The first type, illustrated in scenarios (a-1) and (a-2), occurs under severe occlusion. In these cases, the observable motion history provides insufficient evidence of sudden hazards, such as hidden obstacles or abruptly braking vehicles. Consequently, the model tends to extrapolate the recent motion pattern and produces trajectories close to a constant-velocity profile, rather than capturing the emergency braking behavior in the ground truth. The second type, shown in scenarios (b-1) and (b-2), arises at open intersections with sparse surrounding traffic context and without explicit traffic signal information. In scenario (b-1), the ground truth shows the vehicle stopping at the stop line, whereas the model predicts continued forward motion. In scenario (b-2), the target vehicle stops behind another vehicle at the intersection, while the model fails to anticipate the stopping intention of nearby vehicles under the traffic signal. In both cases, the limited availability of informative interactions makes the underlying driving decision difficult to infer from motion history and local map cues alone.

These examples indicate that extremely long-tail failures are often associated with missing or weakly observable safety-critical context, rather than ordinary motion variation. Under such conditions, even a strong history-based predictor may still make incorrect forecasts. A natural direction for future work is to incorporate Vehicle-to-Everything (V2X) communication priors. Vehicle-to-Vehicle (V2V) communication may help recover occluded interactions through shared perception, while Vehicle-to-Infrastructure (V2I) communication may provide real-time traffic signal cues for disambiguating behaviors at signal-controlled intersections. Such information could complement onboard observations and improve robustness in extreme cases.

\section{Conclusion} \label{Conclusion}
This paper introduces SAIL, a novel framework designed for robust long-tail trajectory prediction by deconstructing the problem across multiple attribute dimensions. The process begins with an Attribute-Guided Trajectory Augmentation strategy and an Attribute Feature Extractor, which form a foundational module to enrich rare samples and encode rich, multi-dimensional long-tail attributes. Building upon this, an Adaptive Momentum Contrastive Learning module employs a continuous cosine momentum schedule and similarity-weighted hard-negative mining to learn a highly discriminative unsupervised feature space. Subsequently, our Evolving Feature Clustering strategy provides dynamic, high-quality pseudo-labels that adapt to the evolving feature manifold during training. Finally, a Focused Decoupled Contrastive Learning module utilizes these pseudo-labels and a novel focusing mechanism to refine cluster compactness, paying special attention to hard-positive samples within tail-end classes. These components collectively enable SAIL to systematically address the long-tail challenge by modeling trajectories based on their intrinsic geometric, dynamic, and risk-based attributes.

Extensive experiments on the nuScenes and ETH/UCY benchmarks demonstrate the effectiveness of SAIL. The proposed method achieves strong performance not only on the full dataset, but also on multiple long-tail subsets and model-agnostic worst-case evaluations. In particular, SAIL consistently outperforms strong baselines on challenging nuScenes scenarios while maintaining favorable inference efficiency, highlighting its practical value for real-world deployment. The consistent gains across both vehicle and pedestrian benchmarks further verify its generalization capability. At the same time, the failure case analysis shows that extreme long-tail errors may still arise when critical safety cues are missing or weakly observable, such as under severe occlusion or at signal-controlled intersections with limited contextual evidence. This suggests that future work may benefit from incorporating richer external priors, such as Vehicle-to-Everything (V2X) communication, to further improve robustness in such challenging scenarios.

\printcredits

\section*{Acknowledgements}
This work was supported by the Science and Technology Development Fund of Macau [0122/2024/RIB2, 0215/2024/AGJ, 001/2024/SKL], the Research Services and Knowledge Transfer Office, University of Macau [SRG2023-00037-IOTSC, MYRG-GRG2024-00284-IOTSC], the Shenzhen-Hong Kong-Macau Science and Technology Program Category C [SGDX20230821095159012], National Natural Science Foundation of China [Grants 52572354], the State Key Lab of Intelligent Transportation System [2024-B001], and the Jiangsu Provincial Science and Technology Program [BZ2024055].

\bibliographystyle{model1-num-names}
\bibliography{cas-refs}

\end{sloppypar}
\end{document}